\documentclass[conference]{IEEEtran}

\IEEEoverridecommandlockouts
\usepackage{cite}
\usepackage{amsmath,amssymb,amsfonts}
\usepackage{algorithmic}
\usepackage{graphicx}
\usepackage{textcomp}
\usepackage{xcolor}
\usepackage{multirow}
\usepackage{tikz}
\usetikzlibrary{calc}
\usepackage{hyperref}

\renewcommand{\phi}{\varphi}

\newcommand{\argmax}{\mathop{\rm arg\,max}\limits}

\def\BibTeX{{\rm B\kern-.05em{\sc i\kern-.025em b}\kern-.08em
    T\kern-.1667em\lower.7ex\hbox{E}\kern-.125emX}}
\begin{document}

\title{Docmarking: Real-Time Screen-Cam Robust Document Image Watermarking}

% \and
% \and
% \and
% \and
% \and

\author{
    \IEEEauthorblockN{Aleksey Yakushev\IEEEauthorrefmark{1}, Yury Markin\IEEEauthorrefmark{1}, Dmitry Obydenkov\IEEEauthorrefmark{1}, Alexander Frolov\IEEEauthorrefmark{1}, Stas Fomin\IEEEauthorrefmark{1}, \\ Manuk Akopyan\IEEEauthorrefmark{1}, Alexander Kozachok\IEEEauthorrefmark{2} and Arthur Gaynov\IEEEauthorrefmark{3}}
    \IEEEauthorblockA{\IEEEauthorrefmark{1}Ivannikov Institute for System Programming of the RAS
    \\\{yakushev, ustas, obydenkov, aefrolov, fomin, manuk\}@ispras.ru
    \\Moscow, Russia}
    \IEEEauthorblockA{\IEEEauthorrefmark{2}Russian Federation Security Guard Service Federal Academy
    \\a.kozachok@academ.msk.rsnet.ru
    \\Oryol, Russia}
    \IEEEauthorblockA{\IEEEauthorrefmark{3}Ministry of Defence of the Russian Federation
    \\gae@mil.ru
    \\Moscow, Russia}
}

\maketitle

% \IEEEpubid{979-8-3503-9853-3/22/\$31.00~\copyright~2022 IEEE}

% \thispagestyle{plain}
% \pagestyle{plain}
% \setcounter{page}{142}

\begin{tikzpicture}[remember picture, overlay]
\node at ($(current page.south) + (0,0.65in)$) {
\begin{minipage}{\textwidth} \footnotesize
  Yakushev A., Markin Yu., Obydenkov D., Frolov A., Fomin S., Akopyan M., Kozachok A., Gaynov A. Docmarking: Real-Time Screen-Cam Robust Document Image Watermarking. 2022 Ivannikov Ispras Open Conference (ISPRAS), IEEE, 2022, pp. 142-150, DOI: \href{https://doi.org/10.1109/ISPRAS57371.2022.10076265}{10.1109/ISPRAS57371.2022.10076265}.\\
  \textcopyright~2022 IEEE. Personal use of this material is permitted. Permission
  from IEEE must be obtained for all other uses, in any current or future media,
  including reprinting/republishing this material for advertising or promotional
  purposes, creating new collective works, for resale or redistribution to
  servers or lists, or reuse of any copyrighted component of this work in other
  works.
\end{minipage}
};
\end{tikzpicture}

%% Old Abstract
%%This paper focuses on protecting confidential documents from leakage by photographing the monitor screen.The user identifier is embedded in the screen image in the form of an imperceptible watermark. An algorithm that generates an image from the message to be encoded using a neural network is developed. The embedded watermark remains permanent and does not depend on the current screen image. The developed decoding algorithm extracts the message encoded in the image with high accuracy. Based on the proposed algorithms, a software package for watermarking text documents on the monitor screen is implemented. Testing shows that the developed watermarking method protects text documents on monitor screen.
\begin{abstract}
This paper focuses on investigation of confidential documents leaks in the form of screen photographs. Proposed approach does not try to prevent leak in the first place but rather aims to determine source of the leak. Method works by applying on the screen a unique identifying watermark as semi-transparent image that is almost imperceptible for human eyes. Watermark image is static and stays on the screen all the time thus watermark present on every captured photograph of the screen. The key components of the approach are three neural networks. The first network generates an image with embedded message in a way that this image is almost invisible when displayed on the screen. The other two neural networks are used to retrieve embedded message with high accuracy. Developed method was comprehensively tested on different screen and cameras. Test results showed high efficiency of the proposed approach.
\end{abstract}


\section{Introduction}\label{introduction}
With the growth of computing power and the acceleration of data processing, more and more areas of human activity are transferred to the digital space.
In particular, usage of electronic documents is growing.
Many organizations that use electronic documents in their processes face document leakage problem.
Leaked documents may contain confidential information intended only for certain company employees. Their transfer to third parties may cause serious financial and reputational losses.

% есть же за 2021 год отчет
According to the InfoWatch analytical center report on the study of restricted information leaks in Russia in 2020 \cite{infowatch}, 79\% of leaks were provoked by insiders, and 79.3\% of them were committed intentionally.
To prevent insider leaks, Data Leakage Prevention (DLP) systems are implemented.
DLP solutions are software and hardware systems that prohibit (active DLP) or register (passive DLP) the actions of company employees when they are working with certain software.
These actions include: sending confidential files via email, using removable USB drives, sending files to cloud storage, etc.

%However, DLP systems do not protect all leak channels.
%The document displayed on the computer screen can be photographed using the employee's personal device.
%After that the screen photo may become publicly available.
%It is almost impossible to prevent this type of leakage without the use of stringent control measures.
%However, there is an approach to simplify the leak investigation. 
%Additional information containing time, department, employee or device identifier is embedded to the displayed image.
%This information present on the screen photo and can be extracted and recovered during the leak investigation.

%% ^^^^^
%% сверху и снизу примерно одно и то же

However, DLP systems do not cover all leak channels, namely, printed copies and photographs of confidential documents displayed on workstation monitor.
Printed copies may be physically moved out from protected perimeter and scans or photographs of printed documents may be anonymously published or send to outsiders (\textit{print-cam} and \textit{print-scan} scenarios).
Screen photographs can be easily taken, as nowadays everyone has personal smartphone with high quality camera, which may be used to breach sensitive information from the screen (\textit{screen-cam} scenario).

%% Embedding additional information on the image refers to the technology of digital watermarks.
%% Watermark can be visible or invisible.
%% A visible watermark is clearly distinguishable image added to host image (e.g. cover text or copyright watermark).
%% If the goal is to hide from the user the fact of the presence of additional embedded information, invisible watermarks are used.
%% They are almost imperceptible for human eyes, but can be recognized by the extraction algorithm.

This study propose data leakage investigations approach committed by insiders using personal smartphone cameras and workstations.
It is supposed that employee intentionally or occasionally may take a photograph of a monitor with confidential information displayed.
Organization provides workstation with preinstalled software, which imposes imperceptible watermark on whole monitor area.
Each employee is associated with a unique watermark that allows to identify the department, user, device, and time of activity.
If the employee takes a photograph of the screen with sensitive information, the photograph contains watermark.
The spreading of watermarked photographs is undesirable.
After getting the photograph the organization security officer can conduct investigation, extract the watermark, and determine the causer of the leakage.
Since the investigation is conducted \textit{post factum}, human guided processing of the watermarked image and slow decoding algorithms can be applied.
The described problem has become a basis for the development of screen watermarking software solution.

\section{Related Works} 

\subsection{Watermarking Methods Typology}

Early studies in data watermarking were published in ‘90s, basic ideas were developed further and dozens of articles appeared during this period \cite{brassil1995electronic, cox1997secure, hartung1999multimedia}.
Researchers classify watermarking techniques by the type of content: text, image, video or audio data.
Usually watermarking algorithm focuses on specific type of data and works inefficiently on other types.
Text documents watermarking approaches may be divided into frequency or spatial domain.
Discrete Fourier Transform (DFT) \cite{premila_2008} and Discrete Cosine Transform (DCT) \cite{dong_2002} based methods refer to the frequency domain. The spatial domain methods apply document modification using information about document text structure: layout, semantic or syntax language properties, format features, etc. They are subdivided into structural and linguistic methods \cite{ahvanooey_18}.

\subsection{Methods for Watermarking Document Images on the Screen}
%Majority of the existing document watermarking methods are intended to be used in print-scan and print-cam scenarios.
%Not many approaches have been proposed for the screen-cam scenario.
%This section describes three methods that work in screen-cam scenario.

\begin{figure}[h]
\begin{minipage}[t]{0.325\linewidth}
\center{\includegraphics[width=1\linewidth]{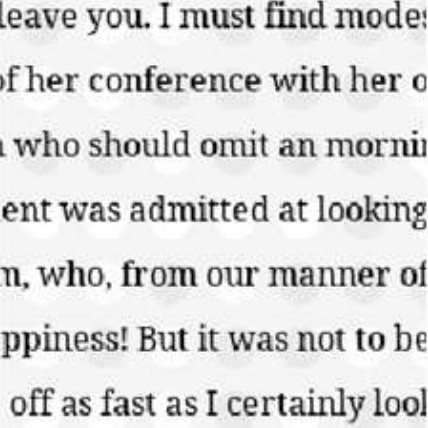}} \\ a) Example of \cite{gslh} marking (taken from \cite{fang2019camera})
\end{minipage}
\hfill
\begin{minipage}[t]{0.325\linewidth}
\center{\includegraphics[width=1\linewidth]{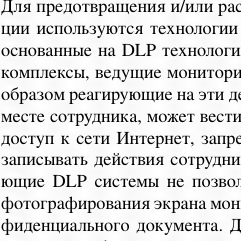}} \\ b) Example of \cite{our_gslh} marking
\end{minipage}
\hfill
\begin{minipage}[t]{0.325\linewidth}
\center{\includegraphics[width=1\linewidth]{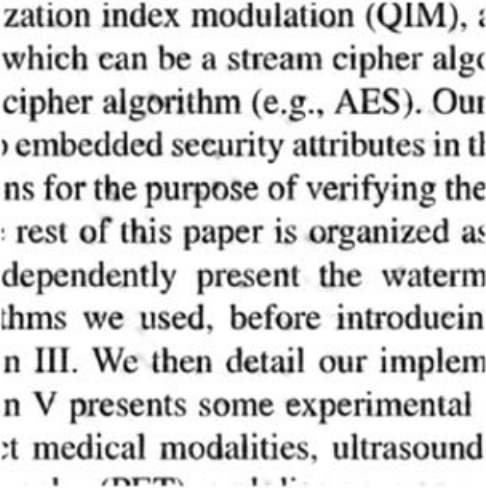}} \\ c) Example of \cite{dnn_screen} marking
\end{minipage}
\caption{Examples of images marked with existing screen-cam robust methods.}
\label{fig:comparison}
\end{figure}

The authors of the paper \cite{gslh} proposed a method for marking text documents based on a smooth change in the brightness of screen image areas.
The watermark is embedded by lowering or brightening the circular areas, depending on the value of the watermark message bits.
A smooth change in brightness is imperceptible to human eyes, but is distinguishable by a digital camera.
Fig.~\ref{fig:comparison}a is an example of a screen photograph with a marked image.
Authors tested the method on edited (scaling, color correction, brightness, contrast, white balance) and unedited screen photographs.
However, the impact of shooting angle and image compression on the extraction accuracy was not studied.

The second approach \cite{our_gslh}, developed as a part of the document watermarking system \cite{docmarking}, is also based on smooth brightness changes on the screen.
The watermark is embedded by placing a sequence of rectangular areas of various brightness into text line spacing.
Imperceptibility of the watermark is based on the assumption that changes in brightness are less visible between the text lines than on wide plain areas (margins).
Fig. \ref{fig:comparison}b is an example of a marked document.
%Despite good test results, the approach turned out to be not very practical.
To embed the watermark, algorithm performs document line spacing search on the screen image.
Delay for rendering the watermark image reaches a few hundreds milliseconds, so the approach causes visual discomfort for users due to insufficient UI responsiveness.
%Such a delay is small enough for the watermark to match the text on the screen photograph, but at the same time causes visual discomfort for the device user.

Under the other approach watermarking of documents images displayed on the screen is implemented by using neural network \cite{dnn_screen}.
The watermark on document image is rather imperceptible (Fig.~\ref{fig:comparison}c); the extraction neural network provides high accuracy.
The authors of the article compared the proposed method with common images watermarking methods robust to screen-cam process.
The comparison showed a significant advantage of the new approach in document image watermarking.
Watermarking of all documents in real time mode on the screen requires frequent launch of embedding neural network which may need high processing speed.
It seems that practical usage of the approach may cause significant difficulties.

%Practical usage of the method remains questionable.

In the work \cite{fang2019camera} the watermark is embedded as a visible noisy template on text background, referred to as underground.
The watermark is designed not to be imperceptible, but inconspicuous, that means it doesn't affect readability of the text and that it can't be extracted without specific algorithm.
As the watermark is visible, the approach doesn't meet the requirement of watermark imperceptibility.

In \cite{fang2021tera} quite similar to \cite{gslh} idea of brightness symbols is used, but imperceptibility is reached in a different way.
If the refresh rate of screen image is high enough, two subsequent frames displayed on the screen may superimpose for human perception, but still can be distinguished by digital camera.
Two subsequent frames contain opposite watermark images, so that their superposition looks like not marked image.
Nevertheless, digital camera captures one of the two frames, so that the photograph of the screen contains watermark.
The drawback of this approach is that it may show low visual quality on most common monitors with the refresh rate of 60 Hz or less.

The usage of moir\'{e} pattern for watermark embedding is proposed in \cite{cheng2021mid} (see \ref{robustness} for more information about moir\'{e} pattern).
Special template is drawn on screen image, that causes moir\'{e} pattern appearance on the captured screen photograph.
The template specifies the form of the moir\'{e} pattern, allowing to embed multibit watermark and pass it to camera via moir\'{e} effect.
The method has a serious disadvantage: it works only under restricted shooting conditions including low range of distances and angles between camera and screen.
Also it has rather low embedding capacity of 14 bits.

All reviewed studies have serious drawbacks, that may face difficulties with implementation of a software solution that fit the previously mentioned problem statement.

% The review of existing solutions showed the need to develop a new algorithm for marking text documents on a monitor screen.
% The main requirement for the new approach is --- it must be static, that is, the watermark image must not change during operation.
% Due to this requirement, the method does not irritate device users and consumes minimal system computing resources.

\section{Proposed Approach}
\subsection{Watermark Screen-Shooting Robustness}\label{robustness}

Embedded watermark must be decoded as accurate as possible, but remain as imperceptible as possible.
Likewise watermarking algorithm must be robust to different image distortions caused by conversion from digital to analog domain and vice versa, lossy compression, and other factors.
%% These contradictory requirements force to balance between different criteria.
This leads to the need to consider in detail the nature of distortions that can be divided into groups, depending on the moment of their appearance.

%When developing an image watermarking method, it is necessary to consider the scenario in which it is used.
%The set of transformations and distortions applied to a watermarked image by the time it enters the mark extraction stage depends on the conditions for applying the method.
%A watermark \textcolor{red}{is said to be} robust to an image transformation if it retains the ability to be extracted without errors after applying this transformation to the watermarked image.
%The figure \ref{fig:photo_examples} shows examples of changing the original image when photographing the screen on which it is displayed.

\begin{figure}[h]
\begin{minipage}[t]{0.24\linewidth}
\center{\includegraphics[width=1\linewidth]{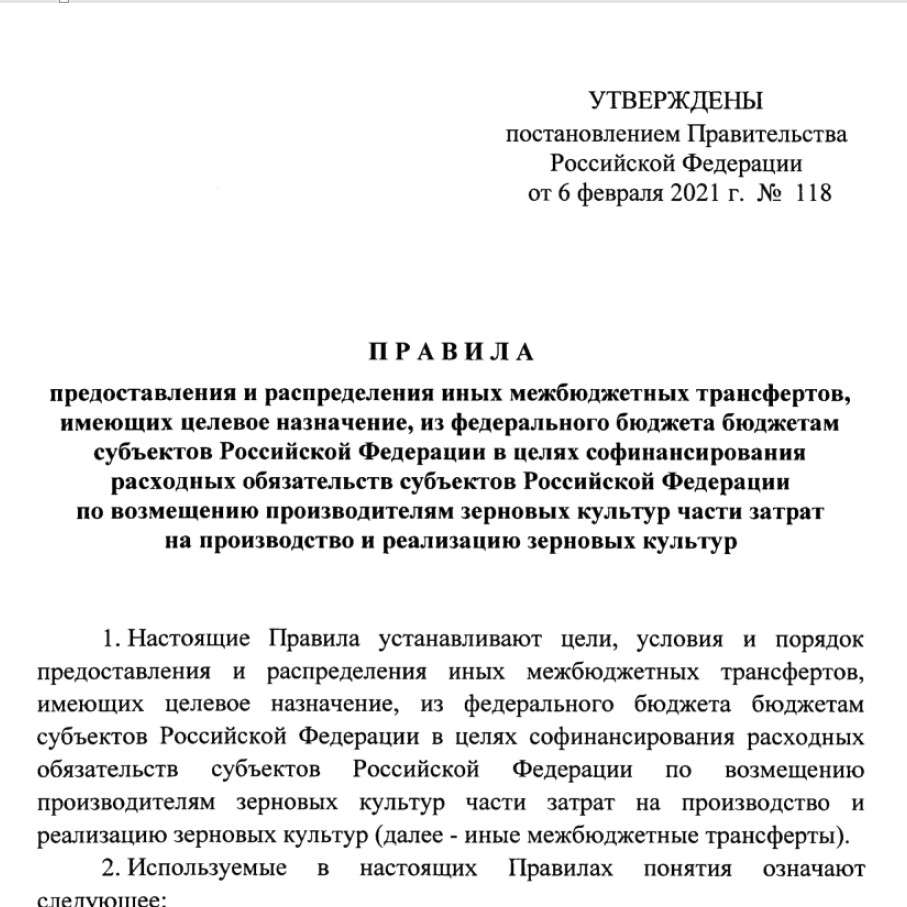}} \\ a) Original document image
\end{minipage}
\hfill
\begin{minipage}[t]{0.24\linewidth}
\center{\includegraphics[width=1\linewidth]{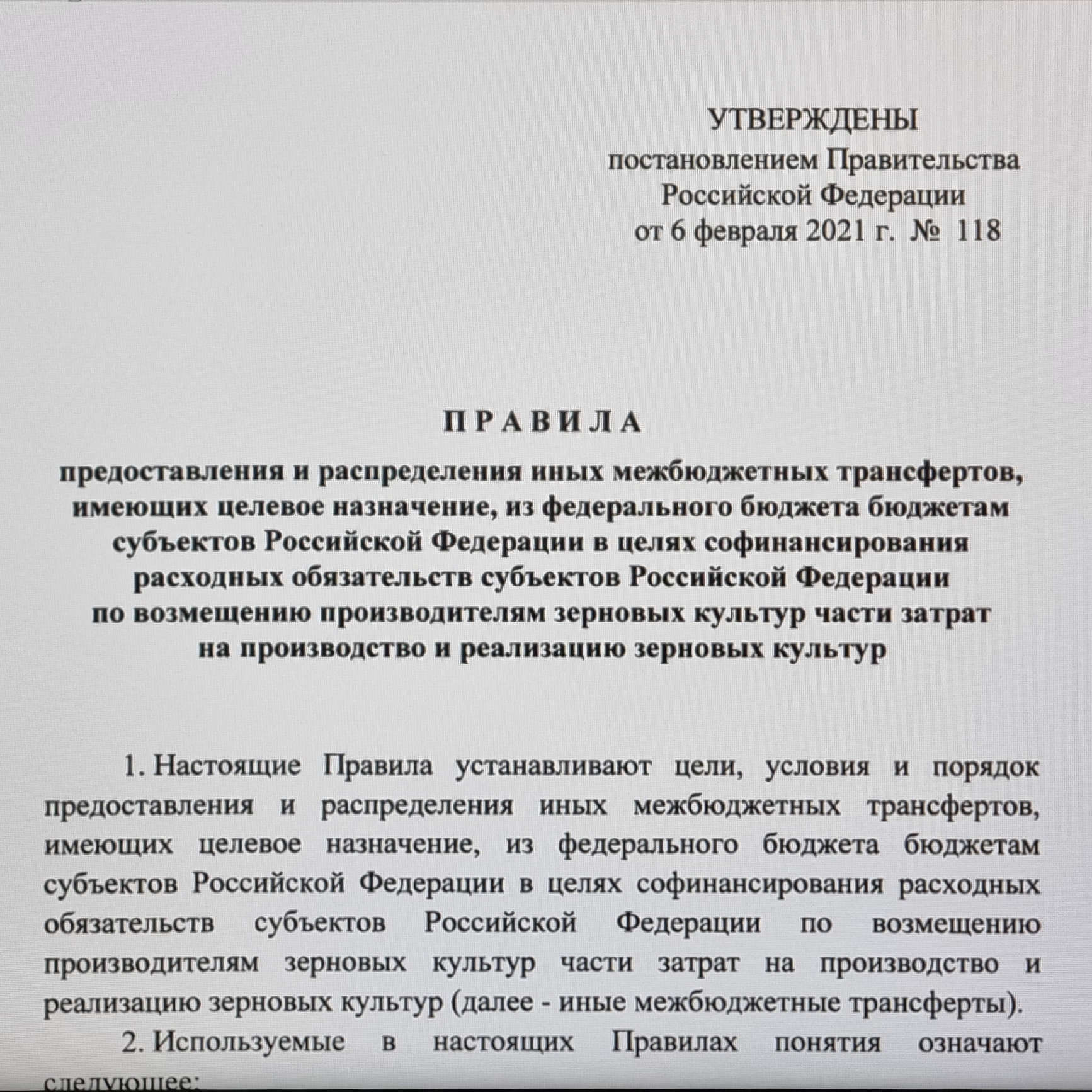}} \\ b) Screen photograph
\end{minipage}
\hfill
\begin{minipage}[t]{0.24\linewidth}
\center{\includegraphics[width=1\linewidth]{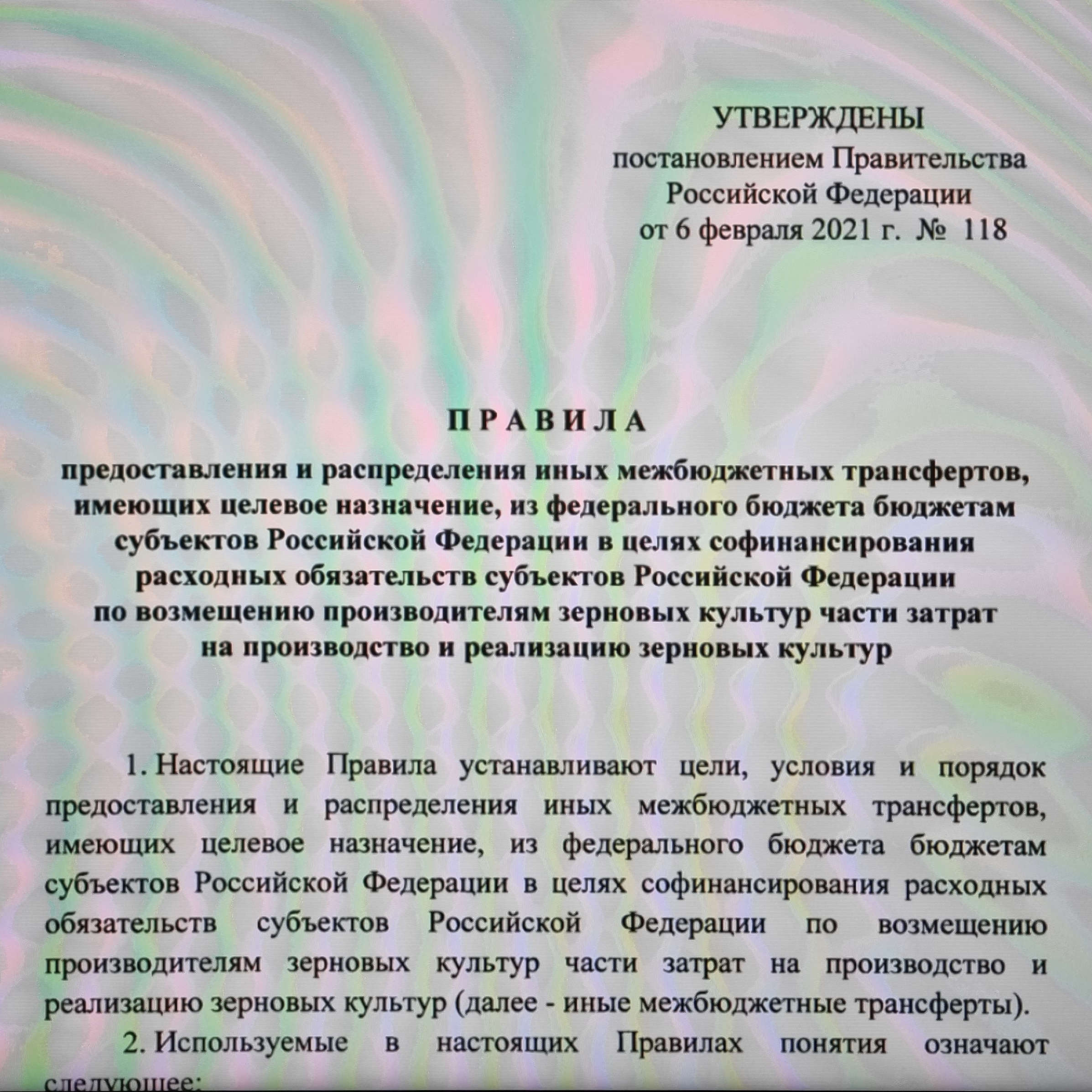}} \\ c) Screen photograph with moir\'{e} effect
\end{minipage}
\hfill
\begin{minipage}[t]{0.24\linewidth}
\center{\includegraphics[width=1\linewidth]{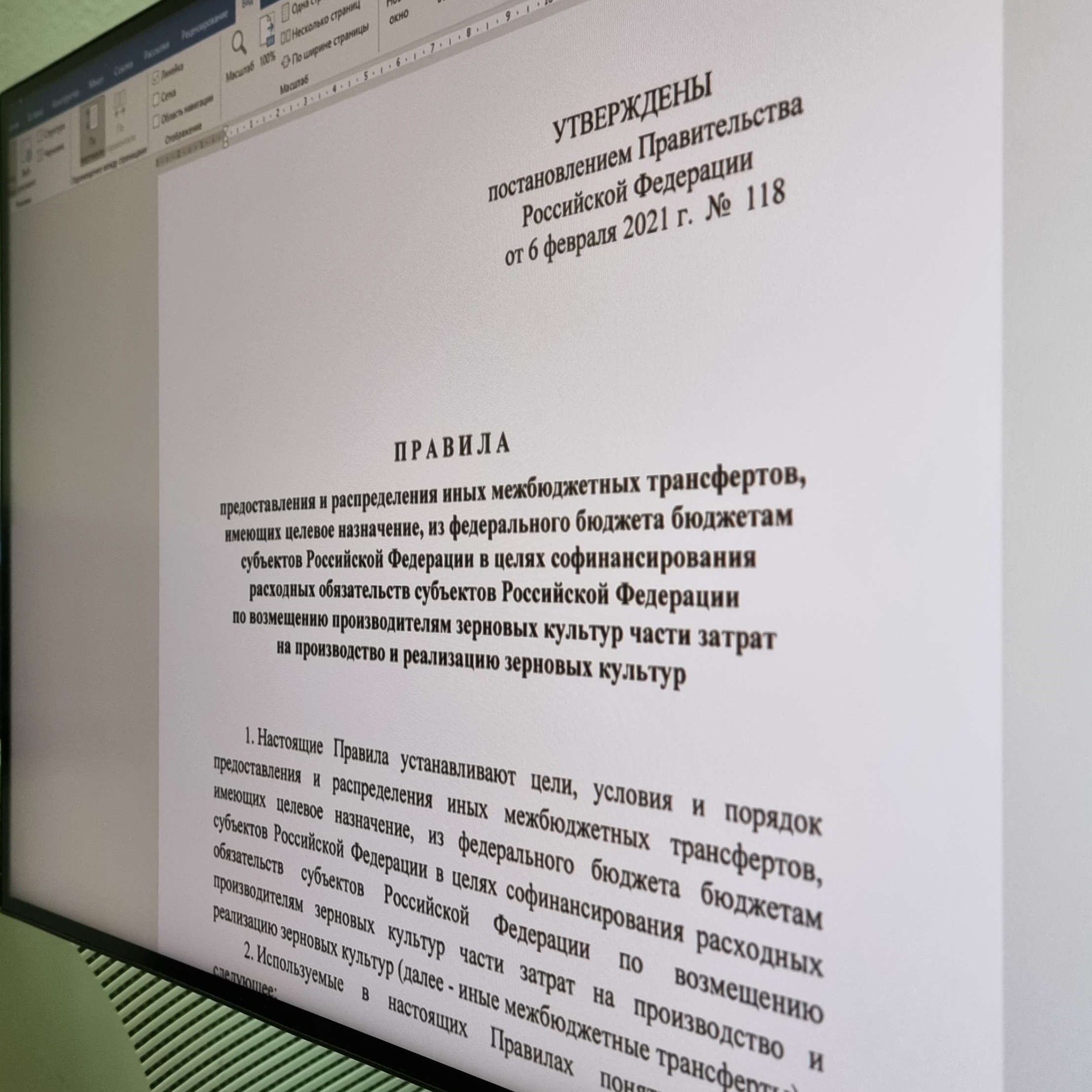}} \\ d) Perspective distorted screen photograph
\end{minipage}
\caption{Photographs of screen displaying document.}
\label{fig:photo_examples}
\end{figure}

The first group of distortions is directly related to the process of displaying an image on the screen.
In this case, the digital signal is converted to analog.
Different monitors significantly vary in their ability to convert the numerical pixel values of an image into the color and luminosity captured by the human eye or a digital camera.
The image is also affected by the characteristics and settings of the monitor.

Distortions from the second group occur at the moment of screen photographing and depend on the photographing conditions.
The camera may be located at different distance and angle from the screen, so the original shape and scale of the image may change.
By the way light is passing thought camera optical system causing different optical effects such as barrel distortion.
Shooting conditions are also affected by the presence of additional light sources and camera focus.

It is important to mention distortion that occurs when photographing a screen --- the moir\'{e} effect.
Usually, the moir\'{e} effect on screen photographs looks like a set of bands of different colors with a non-periodic structure.
The thickness of these bands is not constant and varies in different parts of the photograph.
The moir\'{e} effect occurs when two periodic structures located with some deviation relative to each other are superimposed.
So, when photographing the screen, the pixels of the screen matrix and the sensors of the camera matrix act as such structures.
At the moment, the problem of removing the moir\'{e} effect from photographs to improve their quality is actively studied \cite{moire_challenge}, including screen photographs~\cite{screen_cam_moire}.

%The third group of distortion refers a set of algorithms for processing the resulting photograph.
%The monitor's analog radiation signal is converted into a digital image.
%The sensors of the camera, arranged in the form of a matrix, record the intensity of the light falling on them.
%To register a color image, multi-colored light filters are superimposed on the camera sensors, the location of which is set by the Bayer filter.
%The photograph obtained in this form is subjected to a set of processing algorithms on the device, which can have a significant impact on the embedded watermark.
%After processing, the photo can be additionally subjected to a compression or format conversion procedure.

The third group of distortions refers to a set of algorithms for processing the resulting photograph on the camera device.
Conversion from analog signal domain to digital image significantly differs.
Camera matrices have different light sensitivity, camera manufacturers use various raw photograph enhance algorithms.
And finally, photo is saved into file using lossy compression algorithm.
To sum up, screen cam scenario specific distortions consider serious efforts to achieve acceptable robustness and should be considered while testing.

\subsection{Overview of the Proposed Watermarking Method}
\begin{figure*}
\center{\includegraphics[width=1\linewidth]{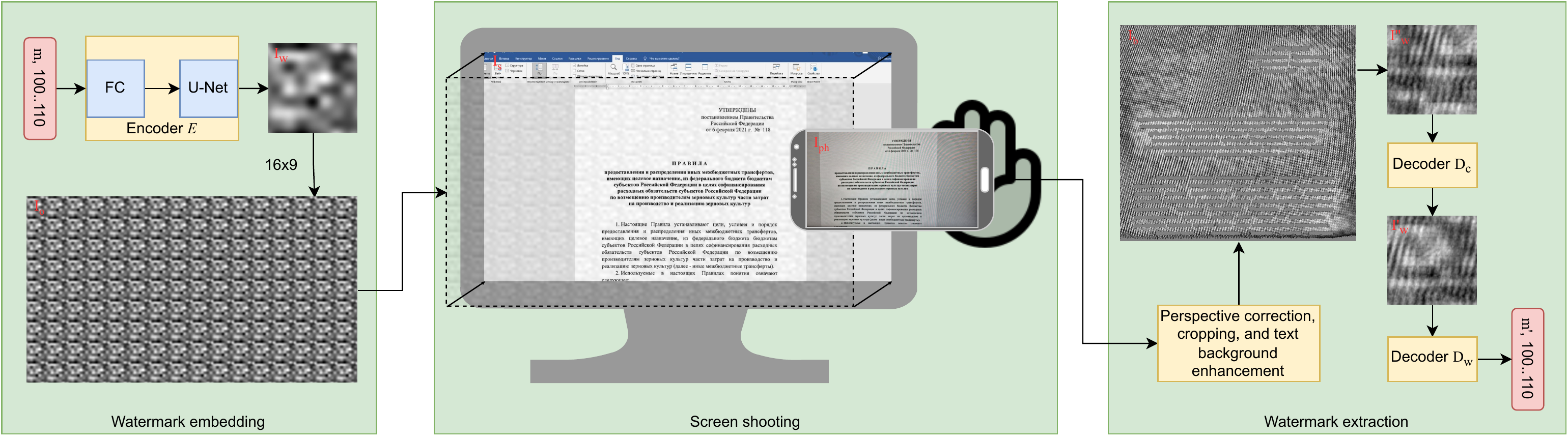}}
\caption{Scheme of the proposed method.}
\label{fig:nss_scheme}
\end{figure*}

The scheme of developed screen watermarking method is shown in Fig. \ref{fig:nss_scheme}.
The encoder neural network $E$ takes a message $m, m_i \in \{0,1\}, i \in \{1, \cdots, M\}$ as input (the method was tested with $M = 50$) and generates a greyscale watermark image $I_w$ with a fixed size $S\times S\times 1$ ($S = 120$ in the implementation). The main properties of image $I_w$ are as follows:
\begin{itemize}
    \item The image $I_w$ consists of smooth brightness transitions;
    \item Smooth brightness transitions are preserved if two identical $I_w$ images are combined, placed either side by side or one above the other.
\end{itemize}
An image $I_o$ is formed by composing several images $I_w$ placed side by side in the form of a grid.
So, for a screen with a resolution of $1920\times1080$ and $S=120$, a grid is composed of $16\times9$ images $I_w$.

The generated image $I_o$ is displayed on the screen with some opacity and covers the screen image $I_s$.
It creates smooth transitions of brightness on the displayed image $I'_s$ due to the described property.

As the encoder network does not use the current screen image $I_s$ to create watermark image, $I_o$ remains static no matter what the user of the device does.
Cover image $I_o$ can be generated once and used through the whole work flow. %%\textcolor{red}{the whole work flow}.
Also, static image $I_o$ does not cause discomfort when working with the device.
The watermark always present on the screen, which allows the method to be used in real-time.

The watermark $m'$ is extracted from the photograph of the screen, with perspective correction and cropping of non-screen areas beforehand.
Due to the periodic structure of the image $I_o$, the watermark presents in all areas of the screen and can be extracted from a photograph of any sufficiently large part of it.
The periodicity of the $I_o$ image is used by watermark extraction algorithm.
Firstly the algorithm determines the value of the period $p$ in the photograph.
Then, an image $I''_w$ is calculated as the average brightness of $p\times p$ areas of the photograph with a $p$ step.
As the top left point of photograph not necessary coincide top left point of screen, the image $I''_w$ may be cyclically shifted relative to $I_w$.
Using the neural network $D_c$, the value of this shift is determined.
The image $I'_w$ is obtained by cyclically shifting the image $I''_w$ by the opposite value.
Then the decoder network $D_w$ is used to determine the bit values of the extracted watermark $m'$.

\subsection{Architecture of the Neural Networks}
The encoder neural network $E$ is used to create an image of a watermark $I_w$ of size $S\times S\times1$ from a message bit sequence of $m$.
It consists of 2 parts.
Firstly an $M$ bit sequence is passed to a fully connected layer in order to obtain a tensor consisting of $S^2$ elements.
The resulting tensor is reshaped to the $S\times S\times1$ format.
It can be interpreted as some preliminary image.
Secondly the tensor is passed to the main part of the encoder network $E$ with close to U-Net\cite{unet} architecture.

An important difference between neural network $E$ and the classic U-Net architecture is the usage of circular padding.
It determines the behavior of convolutional layers on the tensor borders.
The tensor border is processed as if there is a copy of this tensor in the continuation of this border.
Circular padding is used, for example, in the problem of $360^{\circ}$ photographs processing \cite{circular_padding}.
The neural network that receives such images shall take into account that the areas on the left and right borders of the photograph smoothly transit into each other, and this is accomplished by applying circular padding along the horizontal axis.
In the encoder network $E$ this technique is used both along the horizontal and vertical axes.
The main goal of circular padding usage is to make brightness transition on the image $I_o$ as smooth as possible.

In the process of watermark extraction, the image $I''_w$ may be cyclically shifted relative to the image $I_w$.
The neural network $D_c$ is used to find the value of the cyclic shift.
The input of the network $D_c$ is the $S\times S\times1$ image $I''_w$.
The network $D_c$ has the same U-Net architecture as the second part of the neural network $E$, including circular padding.
If the input $D_c$ is the watermark image $I_w$, the tensor $I_c$ of size $S\times S\times1$ is obtained, filled with the following values:

\begin{equation}
    I_c(x, y) = \begin{cases}
    1, & \frac{S}{2} - c \leqslant x, y \leqslant \frac{S}{2} + c \\
    -1, & \begin{split}(0 \leqslant x \leqslant c \text{ or } & S - c \leqslant x \leqslant S) \text{ and }\\ (0 \leqslant y \leqslant c \text { or } & S - c \leqslant y \leqslant S)\end{split} \\
    0, & \text{otherwise} \\
    \end{cases}
\end{equation}
where $c$ --- some given size of center area.

\begin{figure}[t]
\center{\includegraphics[width=0.8\linewidth]{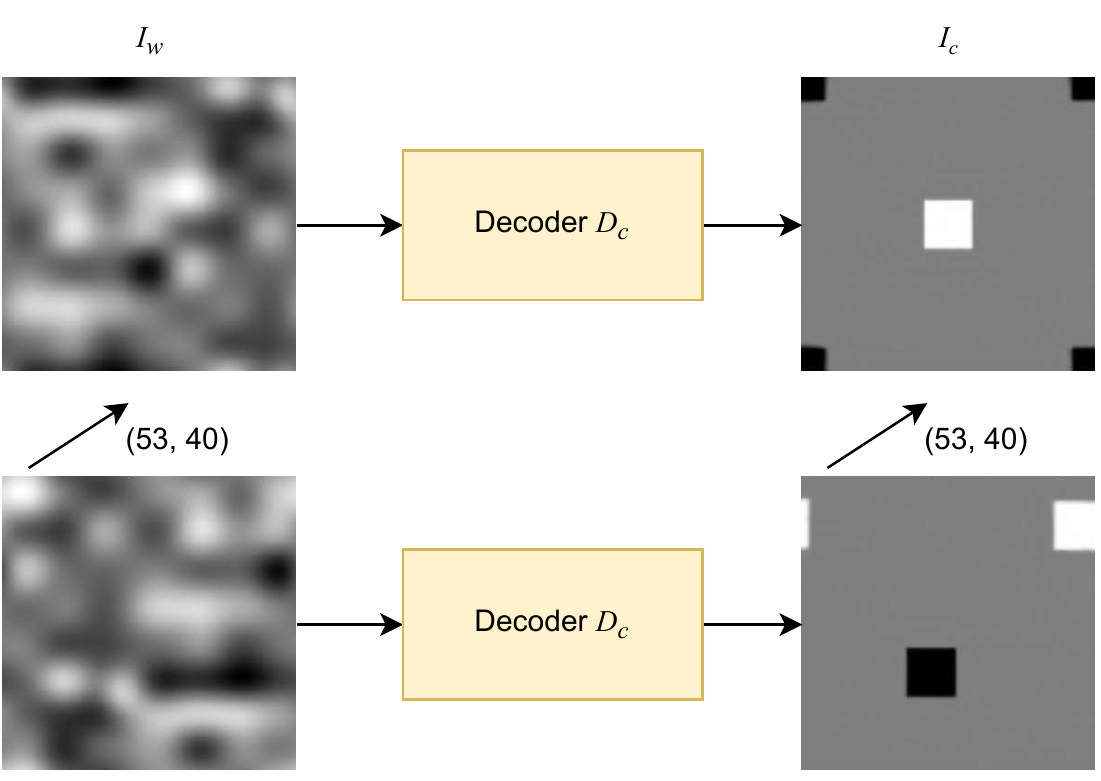}}
\caption{ Decoder $D_c$ invariance to the cyclic shift of the input image.}
\label{fig:d_c_inv}
\end{figure}

The main property of $D_c$ is the invariance to the cyclic shift %%По идее инвариантность значит, что результат не зависит от какого-то фактора, а тут такого нет. Сеть по идее сохраняет значение сдвига, но я хз как это назвать 
of the input image (Fig. \ref{fig:d_c_inv}).
So, if $D_c$ is applied to the image $I_w$, cyclically shifted by some values $(\Delta{x}, \Delta{ y})$, the output will be $I_c$ tensor shifted by the same values. The shift values can be determined by searching for the position of the maximum of the output tensor relative to its center. %%\textcolor{red}{argmax}
In the process of watermark extraction, $D_c$ is used to determine value of the cyclic shift of the image $I''_w$ in order to obtain the image $I'_w$, which differs from $I_w$ by noise caused by the screen shooting process.

The problem of extracting the watermark $w'$ from the image $I'_w$ can be interpreted as a multilabel image classification problem.
Thus, if the bit of the watermark $w_i$ is equal to $1$, it means that image belongs to the class with label $i$. 
Neural network architectures specialized on classification problems can be used as extraction neural network $D_w$.
In the implementation of the proposed method, the EfficientNet-B2\cite{EfficientNet} architecture has been chosen.
The EfficientNet architecture class was obtained using the NAS (Neural Architecture Search) method.
With the same number of trainable parameters, EfficientNet neural networks show the highest accuracy compared to other neural network architectures in the ImageNet\cite{ImageNet} classification problem.

\subsection{Neural Networks Training} \label{training}

\begin{figure}[t]
\center{\includegraphics[width=1\linewidth]{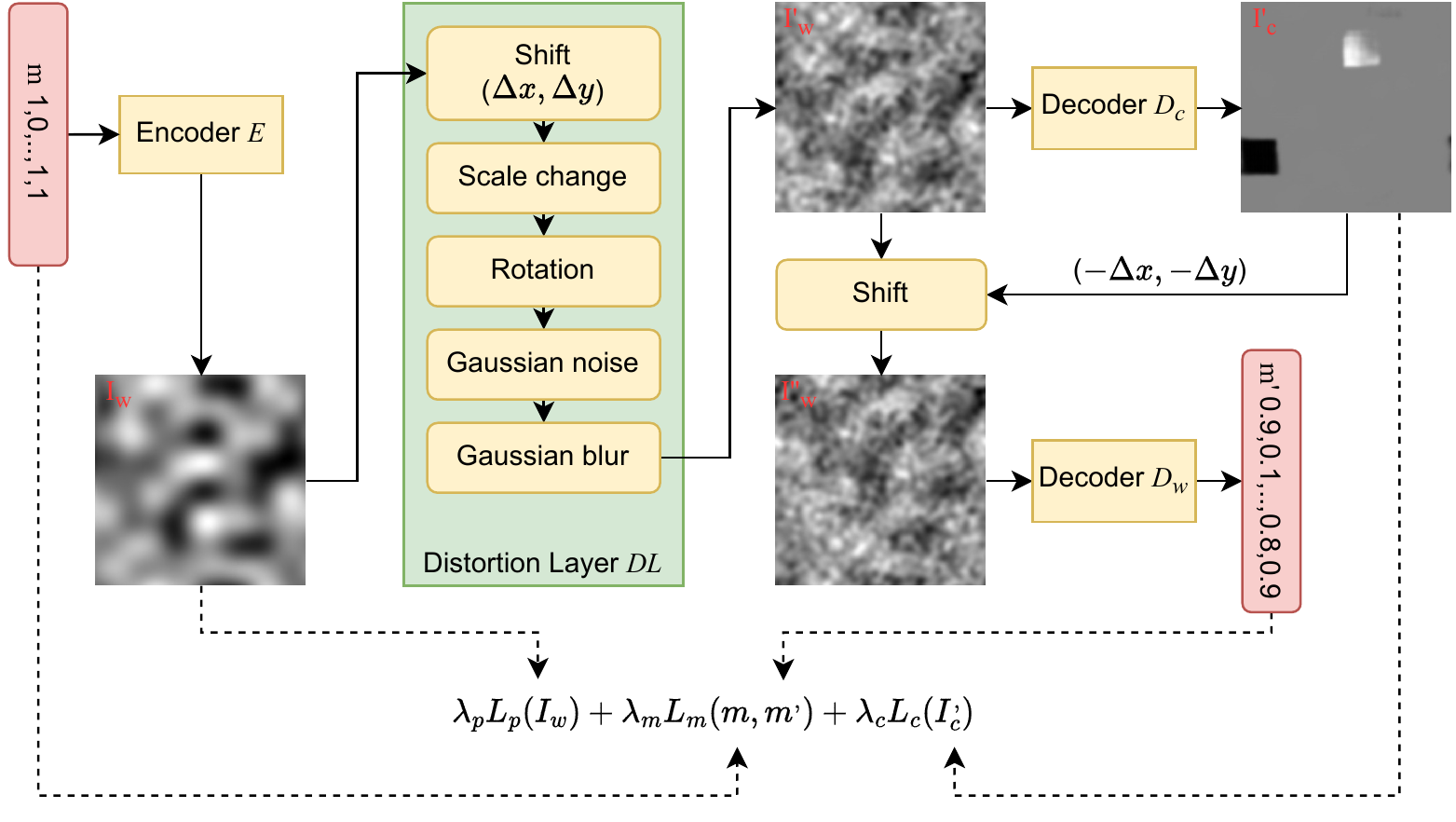}}
\caption{Neural networks training scheme.}
\label{fig:training}
\end{figure}

The neural networks $E$, $D_c$ and $D_w$ are trained simultaneously (Fig. \ref{fig:training}).
Each training iteration consists of the following steps.
First, a random sequence of $M$ bits is generated that defines message $m$.
Based on the message $m$, the neural network $E$ creates an image of the watermark $I_w$.
The image $I_w$ is transformed by the distorting layer $DL$.
The resulting image $I''_w$ is passed to the neural network $D_c$ returning the tensor $I'_c$.
The value of the cyclic shift $(\Delta{x}, \Delta{y})$ of the image $I''_w$ is determined from the position of the $I'_c$ tensor maximum.
The image $I'_w$ is formed by a reverse shift by $(-\Delta{x}, -\Delta{y})$ of the image $I''_w$.
The neural network $D_w$ takes the image $I''_w$ to obtain the extracted $M$ bit sequence $m'$.
Then the loss function $L$ is calculated.
The loss function gradient is used to update the parameters of the neural networks $E$, $D_c$ and $D_w$.

The distortion layer $DL$ is used to make changes to the image $I_w$.
Its main task is to simulate the transformations that occur when photographing the screen during neural networks training.
$DL$ consists of the following steps, performed sequentially:
\begin{enumerate}
    \item Random shift by $(\Delta{x}, \Delta{y})$;
    \item Random scale change within ($0.96$; $1.04$);
    \item Rotation by a random angle within $(-2^{\circ}; 2^{\circ})$;
    \item Applying Gaussian noise $\mathcal{N}(0,1)$;
    \item Applying Gaussian blur with variance $1$.
\end{enumerate}

The loss function $L$ used in the learning process consists of three parts:
\begin{equation}
    L = \lambda_pL_p + \lambda_cL_c + \lambda_mL_m,
\end{equation}
where $\lambda_p$, $\lambda_c$, $\lambda_m$ are the weights of loss functions.

Function $L_p$ specifies the property of a smooth brightness transition on the image $I_w$.
Each pixel of image $I_w$ is compared with its neighbors in a window of size $3\times3$:
\begin{equation}
    L_p = \sqrt{\frac{1}{9S^2}\sum_{\substack{x=1..S\\ y=1..S\\ \delta{x} \in \{-1, 0, 1\} \\ \delta{y} \in \{-1, 0, 1\} }}\left(I_w(x, y) - I_w(x+\delta{x}, y + \delta{y})\right)^2},
\end{equation}
with circular transition on borders:
\begin{equation} \label{borders}
    \begin{split}
        x \notin \{1, \cdots, S\} \Rightarrow x := ((x - 1) \mod S) + 1, \\
        y \notin \{1, \cdots, S\} \Rightarrow y := ((y - 1) \mod S) + 1.
    \end{split}
\end{equation}
Due to the conditions \ref{borders}, the smooth brightness transition on the image $I_w$ has the property of cyclicity.

The function $L_c$ is used to train neural network $D_c$ to determine shift value $(\Delta{x}, \Delta{y})$.
To do this, the $I_c$ tensor is cyclically shifted by $(\Delta{x}, \Delta{y})$, taking into account conditions~\ref{borders}:

\begin{equation}
    I_c^{shifted}(x, y) = I_c (x - \Delta{x}, y - \Delta{y} ).
\end{equation}
The $I'_c$ tensor is compared with the resulting $I_c^{shifted}$ tensor:
\begin{equation}
    L_c = \sqrt{\frac{1}{S^2}\sum_{\substack{x=1..S\\ y=1..S}}\left(I'_c(x, y) - I_c^{shifted}(x, y)\right)^2}.
\end{equation}

The function $L_w$ is responsible for the accuracy of the extracted message.
As mentioned, the task of watermark extracting from an image is similar to an image classification problem, so the binary cross-entropy function can be used to compare the bits of the embedded message $m$ and the extracted message $m'$:
\begin{equation}
    L_w = - \frac{1}{M}\sum_{i=1}^M m_i \log{m'_i} + (1 - m_i) \log{(1 - m'_i) }.
\end{equation}
\subsection{Embedding the Watermark Image on the Screen} \label{imperceptibility}
\begin{figure}[b]
\begin{minipage}[t]{0.325\linewidth}
\center{\includegraphics[width=1\linewidth]{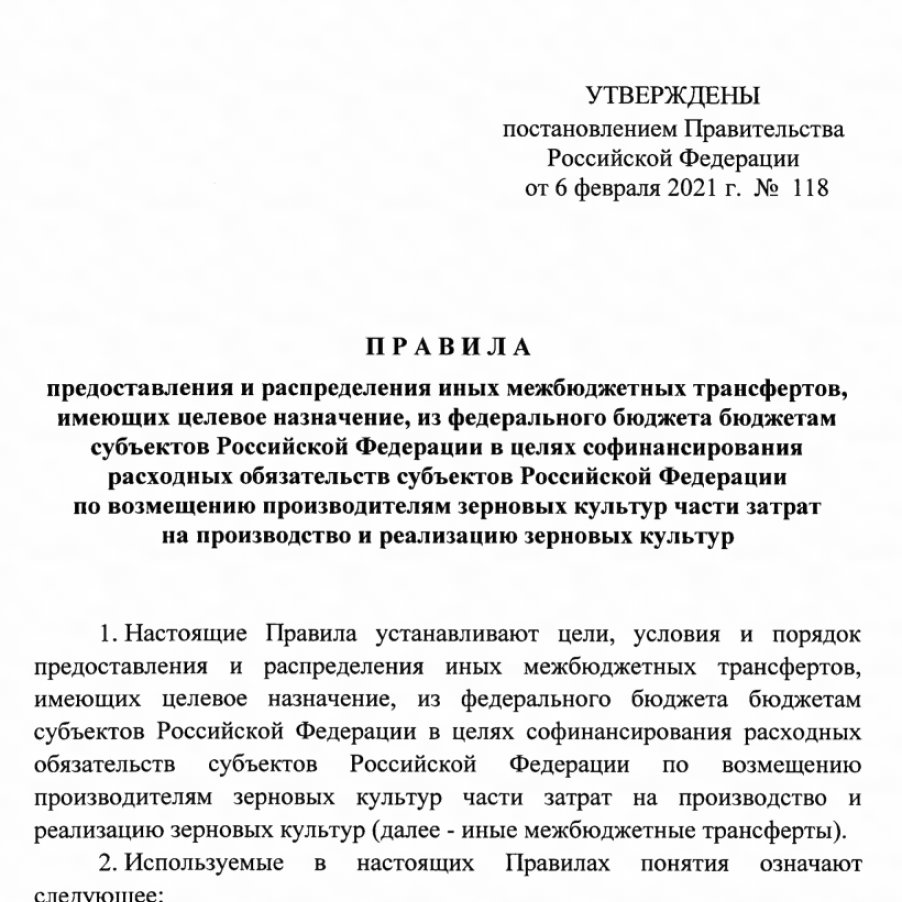}} \\ a) $\alpha = \frac{3}{255}$
\end{minipage}
\hfill
\begin{minipage}[t]{0.325\linewidth}
\center{\includegraphics[width=1\linewidth]{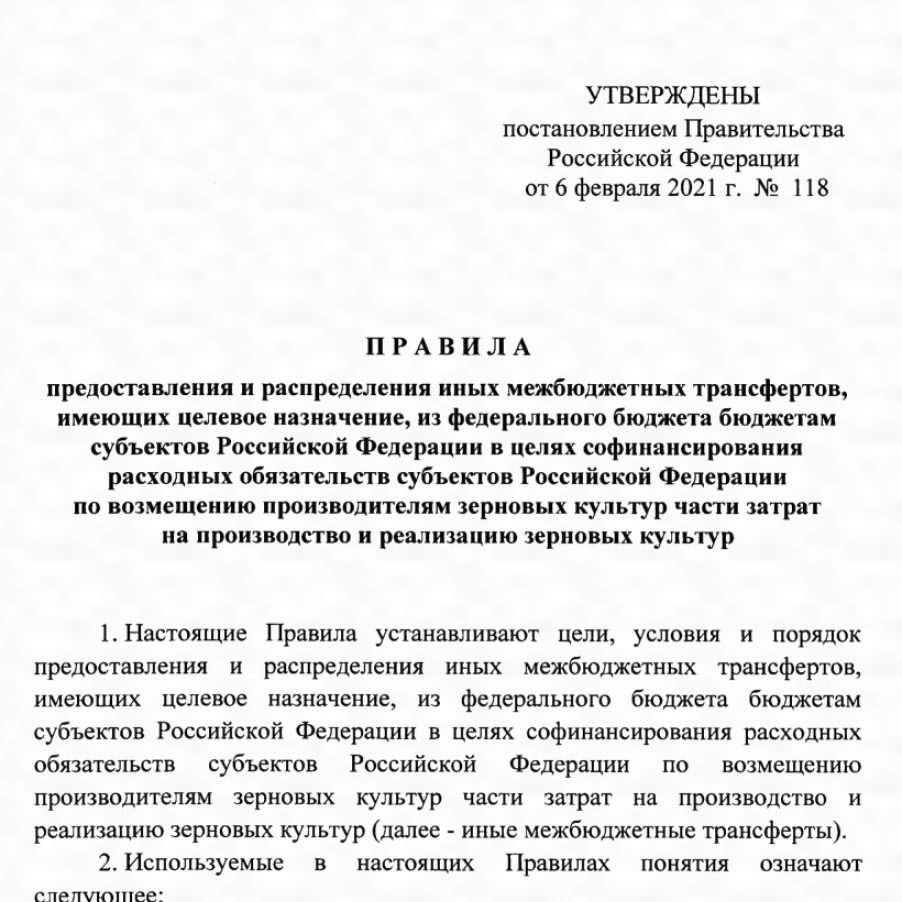}} \\ b) $\alpha = \frac{7}{255}$
\end{minipage}
\hfill
\begin{minipage}[t]{0.325\linewidth}
\center{\includegraphics[width=1\linewidth]{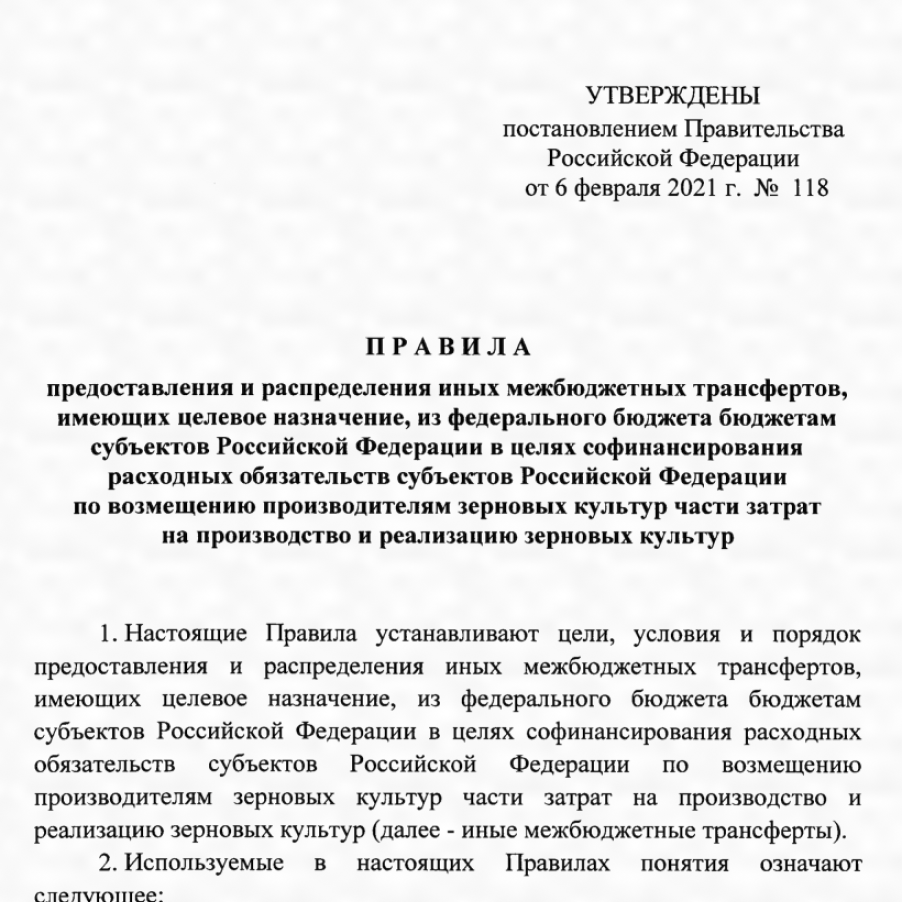}} \\ c) $\alpha = \frac{10}{255}$
\end{minipage}
\caption{Examples of watermarked document images with different opacity $\alpha$.}
\label{fig:nss_example}
\end{figure}

To generate and embed watermark image special software is used.
It is installed on the employee device of the organization that uses the proposed method to protect text documents on the screen. %%\textcolor{red}{employee device} 
The watermarking software interacts with the graphic subsystem of the operating system of the employee's device.
It displays the $I_o$ image containing the watermark in a window, permanently located on the top of all windows in operating system.
The window is fully transparent to user input (keyboard and mouse clicks) and semi-transparent visually with opacity parameter $\alpha \in [ 0, 1 ]$.
According to the rules for adding partially transparent images \cite{compositing}, the pixel values of the image $I'_s$ displayed on the screen are calculated as follows:
\begin{equation}\label{opacity}
   I'^{c}_s(x, y) = (1 - \alpha) \cdot I_s^c(x, y) + \alpha \cdot I_o^c(x, y),
\end{equation}
where $c \in \{R, G, B\}$ denotes the red, green, and blue color channels.
The opacity parameter $\alpha$ determines the imperceptibility of the watermark on the screen.
Examples of marked images with different values of $\alpha$ are shown in Fig. \ref{fig:nss_example}.

% In this work we present a new document watermarking scheme designed for 
% The main goal of this work is to develop and implement a method for marking images of text documents displayed on a monitor screen.
% In this case, a number of requirements must be met:
%     \item The method must support real-time operation: any document displayed on the screen must contain a digital label at any time;
%     \item The digital label must be invisible to the user of the device;
%     \item The numeric label must be taken from a photograph of the screen on which the text document is displayed;
%     \item Documents of any format (Microsoft Word documents, PDF documents, images of scanned documents, etc.) should be marked, regardless of the programs used to display the contents of the document on the screen;
%     \item The digital label must be correctly extracted from the photograph of the labeled screen;
%     \item The program for inserting a digital label on the screen should consume a low amount of computing resources in order to maintain system speed.

%% Присутствует описание поставновки задачи
\subsection{Watermark Extraction Algorithm}
A photograph of a document on a screen that has become publicly available is used by an organization's security analyst to investigate a leak.
Only the leaked photograph is required for the investigation that makes the proposed approach blind. 
It is assumed that the investigation may take some time, thus
watermark extraction algorithm does not have to be very fast.
Moreover, the watermark extraction can be performed repeatedly --- with different algorithm parameters, and some extraction steps can be performed by the analyst using third-party image processing software.
The watermark extraction is conducted in several steps:
\begin{enumerate}
    \item Perspective correction and cropping of non-screen parts of the photograph (done by the analyst beforehand);
    \item Detection of background areas in the document photograph, getting image $I_b$;
    \item Searching for a periodic structure in the image $I_b$, determining the period $p$;
    \item Averaging $I_b$ with step $p$ up to $p\times p$ image $I_p$, rescaling it to obtain the image $I''_w$ of size $S\times S$;
    \item Cyclic shift of the image $I''_w$ using the neural network $D_c$ resulting in the image $I'_w$;
    \item Extracting the bit values of the message $m'$ from the image $I'_w$ by applying the neural network $D_w$.
\end{enumerate}

In most cases, screen photographs are made at some angle to the screen.
Thus, screen image perspective on the photograph is distorted.
The first step in the watermark extraction is to correct the perspective of the photograph.
It can be performed by an analyst using image editing software.
The corrected photograph is cropped in order to leave only the areas related to the screen in the final image.

The cropped photograph is converted to grayscale denoted as $I_{ph}$.
To determine the background areas in the document in the image $I_{ph}$ (step 2), the value in each pixel is compared to the median value in a window of the given size $a \times a$ centered on that pixel.

Image $I_b$ is calculated as follows:
\begin{equation}
    \begin{split}
        &I_b(x, y) = \\ 
        &\begin{cases}
            I_{ph}(x, y) - I_{ph}^a(x, y), & |I_{ph}(x, y) - I_{ph}^a(x, y) | \leqslant t, \\
            0, & |I_{ph}(x, y) - I_{ph}^a(x, y) | > t,
        \end{cases}
    \end{split}
\end{equation}
where $I_{ph}^a(x, y)$ is the median value of $I_{ph}$ in a window of size $a\times a$ centered at $(x, y)$; $t$ is specified threshold value.

Image of a text document is a set of mostly black characters on plain, often white, background.
Due to this property, displayed image $I'_s$ is close to $I_o$ in the document background areas. 
As $I_o$ is a grid of $I_w$ images, it has a periodic structure with period $S$.
Periodic structure is preserved on photograph document background but it has different period $p$ due to scaling.
The purpose of the third step watermark extraction step is to determine the value of the period $p$ in the image $I_b$.
Denote by $I_p$ the averaging of the image $I_b$ with the step $p$:
\begin{equation}\label{i_p}
    I_p(x,y) = \frac{1}{\lfloor \frac{W}{p} \rfloor\cdot \lfloor \frac{H}{p} \rfloor}\sum_{k, l = 1 }^ { \lfloor \frac{W}{p} \rfloor, \lfloor \frac{H}{p} \rfloor} I_b(k \cdot p + x, l \cdot p + y).
\end{equation}

Period determination is based on the following observation.
If the $p$ value does not correspond to the desired period, the distribution of the image $I_p$ pixel values turns out to be close to random noise.
After removing the noise using the Gaussian filter $G(\cdot)$, the image becomes very monotone. The standard deviation $std(\cdot)$ of this image is close to $0$.
Alternatively, If $p$ coincides with the period, applying the Gaussian filter to $I_p$ preserves the brightness transitions and the standard deviation is higher than zero.
Thus, the period can be determined by searching the maximum of the following:
\begin{equation}
    p = \argmax_{p' \in [p_0,\cdots, p_1]} std(G(I_{p'})).
\end{equation}

The averaging of the image $I_b$ at the fourth step also conducted according to the Eq. \ref{i_p}.
The image $I_p$ obtained by averaging $I_b$ is rescaled from $p \times p$ to $S\times S$, resulting in the image $I''_w$. %%Вот это предложение вообще не понял
The further watermark extraction steps are the same as in processing of the image $I''_w$ during neural networks training described in the \ref{training} section.

\section{Experimental Results and Analysis}
Proposed watermarking method was implemented with watermark capacity $M=50$ bits and watermark image $I_w$ with size $S=120$.
Neural networks $E$, $D_c$ and $D_w$ were trained with the Adam optimizer \cite{adam}.
In total, $1440000$ $50$-bit sequences were generated during the training process.

In the practical application of the proposed watermarking method a 32-bit message is embedded.
18 bits of BCH error correction codes \cite{bch} are added to the message, allowing to correct up to 3 errors in the received 50-bit sequence.
Thus, we can assume that the watermark is extracted correctly if there are no more than 3 errors.

To test the method photographs of monitor screens with displayed images of documents were taken.
The websites of the Ministry of Education and Science of the Russian Federation \cite{minobr} and the Ministry of Finance of the Russian Federation \cite{minfin} were chosen as sources for the document images.
The experiments involved 3 monitors and 3 smartphones.
Monitor characteristics are given in the Table \ref{monitors}, and the smartphone digital cameras characteristics are given in the Table \ref{cameras}.
The experiments were carried out using the default camera application on each smartphone in automatic mode.

\begin{table}[b]
    \caption{Monitor Characteristics}
    \centering
    \begin{tabular}{cccc}
    \hline
    Monitor & Matrix Type & Screen Resolution & Refresh Rate\\
    \hline
        Samsung SM 940FN & TFT PVA & $1280 \times 1024$ & 75Hz\\
        Sony SDM-S75A & TN & $1280 \times 1024$ & 75Hz\\
        Dell U2722D & IPS & $2560\times 1440$ & 60Hz\\
    \hline
    \end{tabular}
    \label{monitors}
    
    \caption{Camera Characteristics}
    \centering
    \begin{tabular}{ccccc}
    \hline
    Smartphone & Resolution & Aperture & Focal Length & Sensor Size \\
    \hline
        Xiaomi Mi A1 & 12MP & f/2.2 & 26mm & 1.25um \\
        Samsung S8 & 12MP & f/1.7 & 26mm & 1.4um \\
        Samsung S21 & 12MP & f/1.8 & 26mm & 1.8um\\
    \hline
    \end{tabular}
    \label{cameras}
\end{table}

\subsection{Selecting the Opacity of the Marking Window}
As mentioned in section \ref{imperceptibility}, the imperceptibility of the watermark on the screen is affected by the value of marking window opacity $\alpha$.
Based on the implementation features, it is convenient to represent the values of $\alpha$ in fractions of the form $n/255$, where $n$ is an integer.
To determine the degree of imperceptibility, the Peak Signal to Noise Ratio (PSNR) and Structural Similarity Metric (SSIM) \cite{ssim} were chosen as imperceptibility metrics.
Metrics were averaged over images of 50 marked documents.
Since it is assumed that the watermark is displayed regardless of the screen image, it is important to make it imperceptible not only on document images, but also on common images. Imperceptibility metrics were also calculated for 50 common images taken from the Open Images V6 dataset \cite {open_images}.
The results of imperceptibility metrics calculating for different values of $\alpha$ are presented in the Table \ref{metrics_table}.
\begin{table}[t]
    \caption{Values of Watermark Imperceptibility Metrics}
    \centering
    \begin{tabular}{c|cc|cc}
    \hline
    \multirow{2}{*} {$\alpha$ } & \multicolumn{2}{c|}{Document images} & \multicolumn{2}{c}{Common images} \\ \cline{2-5}
    & $PSNR$, dB & $SSIM$ & $PSNR$, dB & $SSIM$ \\
    \hline
       $3/255$ & 44.4 & 0.9992 & 47.5 & 0.9972 \\
       $4/255$ & 42.6 & 0.9991 & 45.5 & 0.9958 \\
       $5/255$ & 40.6 & 0.9989 & 43.8 & 0.9942 \\
       $6/255$ & 39.0 & 0.9987 & 42.4 & 0.9926 \\
       $7/255$ & 37.7 & 0.9985 & 41.1 & 0.9910 \\
       $8/255$ & 36.5 & 0.9983 & 40.0 & 0.9894 \\
       $9/255$ & 35.4 & 0.9980 & 39.0 & 0.9878 \\
       $10/255$ & 34.5 & 0.9977 & 38.1 & 0.9863 \\
    \hline
    \end{tabular}
    \label{metrics_table}
\end{table}

Increasing the opacity $\alpha$ of marking window leads to an increase in the visibility of the watermark.
According to the $SSIM$ metric, the method demonstrates high imperceptibility for all $\alpha$ values used in testing.
The $PSNR$ metric shows that the embedded watermark is less visible on common images than on document images.
For values of $\alpha$ not exceeding $5/255$, the watermark is hardly noticeable.

To determine the minimal appropriate opacity value $\alpha$, we checked the accuracy of the watermark extraction.
Shooting was performed with $9$ pairs of cameras and monitors in a given range of $\alpha$ values at the distance of 40 centimeters between camera and monitor.
For each pair and each opacity value, $10$ photographs of documents of various sizes displayed on the monitor screen were taken.
Watermarks were extracted from the photographs and compared to the embedded watermarks.
We calculated Bit Error Rate (BER) --- the average amount of inverted bits in extracted watermark before error correction.
If the value of BER is close to 0, almost all watermarks were extracted correctly.
As we use error correction codes, we also counted number of photographs with no more than 3 errors in the extracted watermark. After error correction, these watermarks correspond to the embedded watermarks.
The results of the experiment are presented in the Table \ref{opacity_table}.

\begin{table}[b!]
    \caption{Accuracy of watermark extraction for various opacity values}
    \centering
    \begin{tabular}{c|cc|cc|cc}
    \hline
    \multirow{3}{*} { $\alpha$ } & \multicolumn{6}{c}{Camera} \\ \cline{2-7}
    & \multicolumn{2}{c|}{Xiaomi Mi A1} & \multicolumn{2}{c|}{Samsung S8} & \multicolumn{2}{c}{Samsung S21} \\ \cline{2 -7}
    & BER & $\leqslant 3$ err. & BER & $\leqslant 3$ err. &BER & $\leqslant 3$ err.\\
    \hline
    \multicolumn{7}{c}{Samsung SyncMaster 940FN Monitor} \\
    \hline
    3/255 & 33.6\% & 3/10 & 18.4\% & 7/10 & 17.6\% & 6/10 \\
    4/255 & 9.1\% & 7/10 & 1.4\% & 10/10 & 6.8\% & 9/10 \\
    5/255 & 10.6\% & 7/10 & 0.6\% & 10/10 & 0.2\% & 10/10 \\
    6/255 & 1.0\% & 10/10 & 0.4\% & 10/10 & 0.2\% & 10/10 \\
    7/255 & 0.6\% & 10/10 & 0.0\% & 10/10 & 0.2\% & 10/10 \\
    8/255 & 0.2\% & 10/10 & 0.0\% & 10/10 & 0.2\% & 10/10 \\
    9/255 & 0.2\% & 10/10 & 0.0\% & 10/10 & 0.0\% & 10/10 \\
    10/255 & 0.0\% & 10/10 & 0.0\% & 10/10 & 0.0\% & 10/10 \\
    \hline
    \multicolumn{7}{c}{Sony SDM-S75A Monitor} \\
    \hline
    3/255 & 44.8\% & 0/10 & 50.2\% & 0/10 & 48.0\% & 1/10 \\
    4/255 & 48.4\% & 0/10 & 14.6\% & 8/10 & 36.8\% & 3/10 \\
    5/255 & 41.0\% & 1/10 & 2.0\% & 10/10 & 34.6\% & 3/10 \\
    6/255 & 41.4\% & 2/10 & 1.4\% & 10/10 & 31.6\% & 4/10 \\
    7/255 & 32.2\% & 4/10 & 1.4\% & 10/10 & 21.6\% & 5/10 \\
    8/255 & 6.0\% & 9/10 & 0.8\% & 10/10 & 20.4\% & 6/10 \\
    9/255 & 6.4\% & 9/10 & 0.4\% & 10/10 & 16.8\% & 7/10 \\
    10/255 & 4.0\% & 9/10 & 0.6\% & 10/10 & 11.2\% & 8/10 \\
    \hline
    \multicolumn{7}{c}{Dell U2722D Monitor} \\
    \hline
    3/255 & 7.8\% & 8/10 & 8.8\% & 8/10 & 14.8\% & 7/10 \\
    4/255 & 1.8\% & 9/10 & 1.4\% & 10/10 & 1.2\% & 10/10 \\
    5/255 & 1.0\% & 10/10 & 1.0\% & 10/10 & 1.0\% & 10/10 \\
    6/255 & 0.8\% & 10/10 & 0.6\% & 10/10 & 0.4\% & 10/10 \\
    7/255 & 0.2\% & 10/10 & 0.2\% & 10/10 & 0.0\% & 10/10 \\
    8/255 & 0.0\% & 10/10 & 0.2\% & 10/10 & 0.0\% & 10/10 \\
    9/255 & 0.0\% & 10/10 & 0.0\% & 10/10 & 0.0\% & 10/10 \\
    10/255 & 0.0\% & 10/10 & 0.2\% & 10/10 & 0.0\% & 10/10 \\
    \hline
    \end{tabular}
    \label{opacity_table}
\end{table}

\begin{figure}[t]
\begin{minipage}[t]{0.49\linewidth}
\center{\includegraphics[width=1\linewidth]{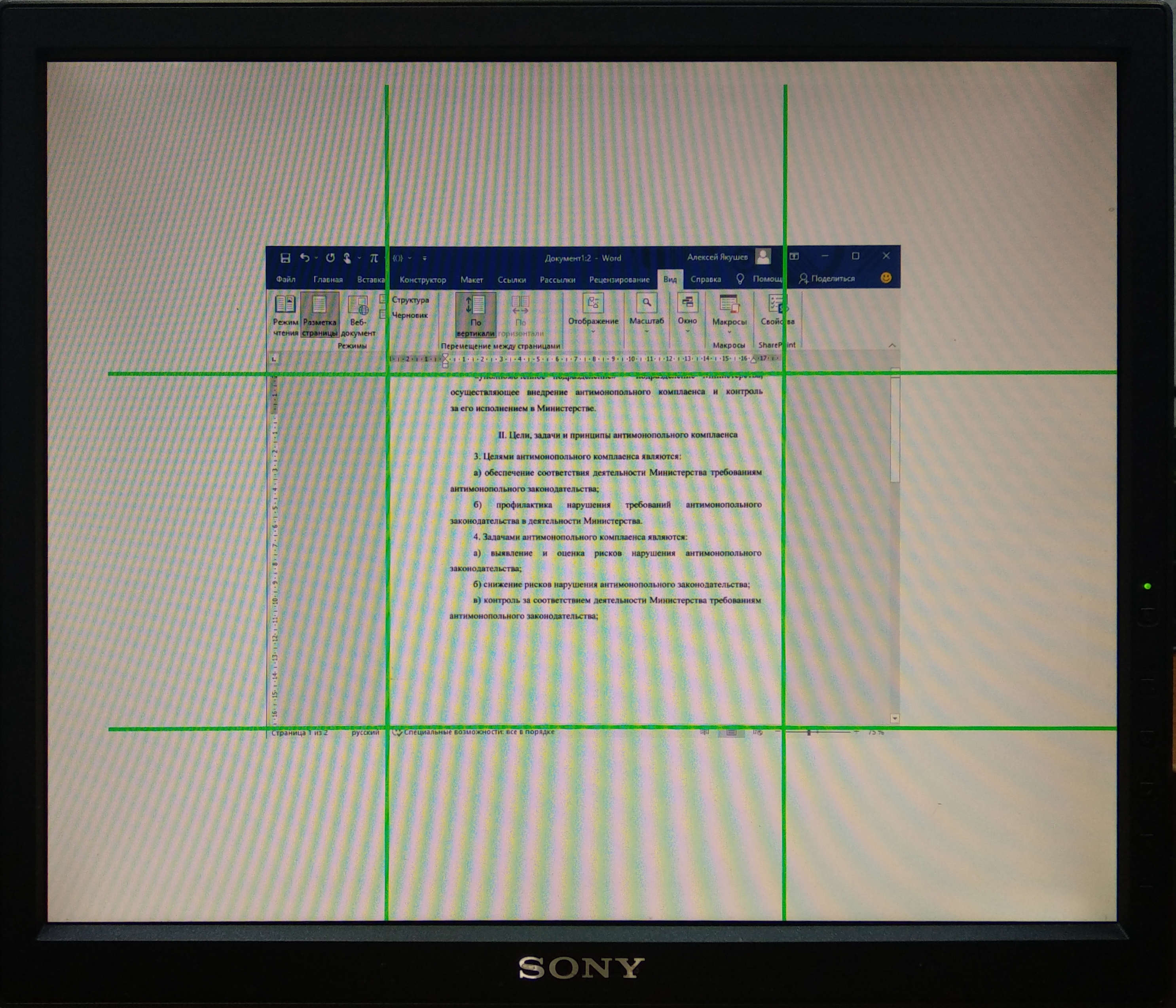}} \\ a) Xiaomi Mi A1
\end{minipage}
\hfill
\begin{minipage}[t]{0.49\linewidth}
\center{\includegraphics[width=1\linewidth]{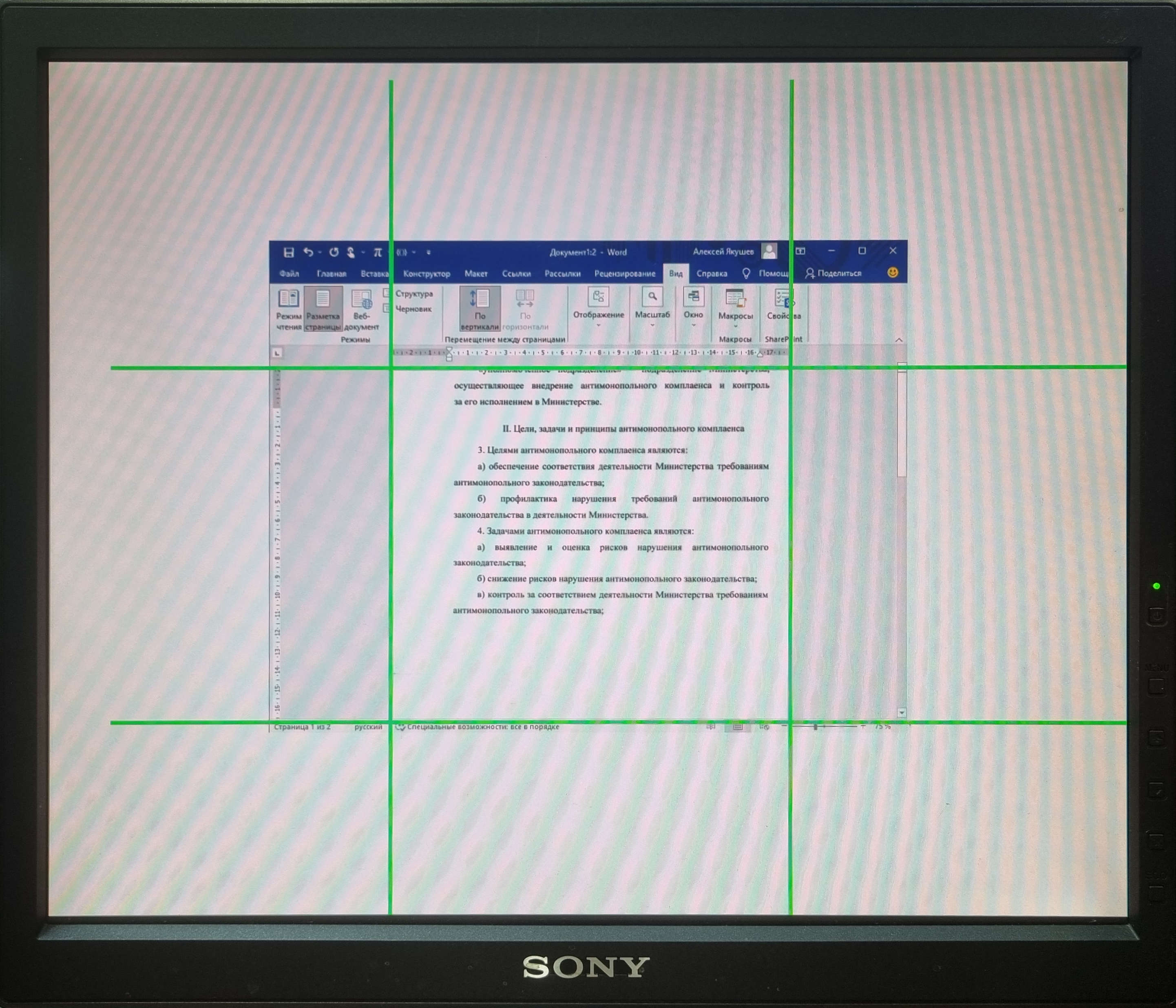}} \\ b) Samsung Galaxy S21
\end{minipage}
\caption{Photograhs of the Sony SDM-S75A monitor at the distance of 40 cm. Green lines on document borders are used for automatic perspective correction and photograph cropping, both can be done manually in practical application.}
\label{fig:moire_photos}
\end{figure}

At $\alpha \geqslant 7/255$, the watermark is correctly extracted from almost all photographs, except for photographs of the Sony SDM-S75A monitor screen taken with the cameras of Xiaomi Mi A1 and Samsung Galaxy S21 smartphones.
These photographs are strongly distorted with the moir\'{e} effect (Fig. \ref{fig:moire_photos}).
In the following experiments, $\alpha$ values were fixed for each monitor: $7/255$ for the Samsung SyncMaster 940FN monitor, $8/255$ for the Sony SDM-S75A monitor, $6/255$ for the Dell U2722D monitor.

\subsection{The Impact of Shooting Distance on the Extraction Accuracy}

For all pairs of monitors and cameras, photographs were taken at different distances between the camera and the screen.
The results of watermark extraction are shown in the Table \ref{distance_table}.
\begin{table}[b]
    \caption{The impact of shooting distance on the watermark extraction accuracy}
    \centering
    \begin{tabular}{c|cc|cc|cc}
    \hline
    & \multicolumn{6}{c}{Camera} \\ \cline{2-7}
    Distance & \multicolumn{2}{c|}{Xiaomi Mi A1} & \multicolumn{2}{c|}{Samsung S8} & \multicolumn{2}{c}{Samsung S21} \\ \cline {2-7}
    & BER & $\leqslant 3$ err. & BER & $\leqslant 3$ err. & BER & $\leqslant 3$ err.\\
    \hline
    \multicolumn{7}{c}{Samsung SyncMaster 940FN Monitor} \\
    \hline
    25 cm & 1.6\% & 10/10 & 45.0\% & 0/10 & 1.4\% & 10/10 \\
    40 cm & 0.4\% & 10/10 & 0.2\% & 10/10 & 0.8\% & 10/10 \\
    60 cm & 0.0\% & 10/10 & 0.8\% & 10/10 & 0.0\% & 10/10 \\
    80 cm & 0.6\% & 10/10 & 1.2\% & 10/10 & 0.2\% & 10/10 \\
    100 cm & 0.4\% & 10/10 & 1.2\% & 9/10 & 0.6\% & 10/10 \\
    \hline
    \multicolumn{7}{c}{Sony SDM-S75A Monitor} \\
    \hline
    25 cm & 34.0\% & 3/10 & 4.8\% & 8/10 & 22.4\% & 6/10 \\
    40 cm & 32.0\% & 4/10 & 0.8\% & 10/10 & 22.0\% & 6/10 \\
    60 cm & 0.6\% & 10/10 & 0.6\% & 10/10 & 0.4\% & 10/10 \\
    80 cm & 0.4\% & 10/10 & 1.2\% & 10/10 & 0.4\% & 10/10 \\
    100 cm & 0.8\% & 10/10 & 2.0\% & 10/10 & 0.2\% & 10/10 \\
    \hline
    \multicolumn{7}{c}{Dell U2722D Monitor} \\
    \hline
    25 cm & 27.8\% & 4/10 & 9.6\% & 5/10 & 30.8\% & 4/10 \\
    40 cm & 1.0\% & 10/10 & 1.2\% & 10/10 & 0.8\% & 10/10 \\
    60 cm & 0.4\% & 10/10 & 7.0\% & 9/10 & 18.6\% & 6/10 \\
    80 cm & 0.2\% & 10/10 & 4.4\% & 9/10 & 0.2\% & 10/10 \\
    100 cm & 0.2\% & 10/10 & 4.4\% & 9/10 & 0.6\% & 10/10 \\
    \hline
    \end{tabular}
    \label{distance_table}
\end{table}

As in the previous experiment, there is a high error percentage in watermarks extracted from photographs of the Sony SDM-S75A monitor screen taken with the cameras of Xiaomi Mi A1 and Samsung Galaxy S21 smartphones at the distance of 40 cm.
Meanwhile, the extraction accuracy at other distances is much higher.
This happens due to the fact that increased moir\'{e} effect in photography occurs under certain shooting conditions, including the distance between the camera and the screen.
Also, on other pairs of monitors and cameras, the moir\'{e} effect appears at the distance of 25 cm.

\subsection{The Impact of Shooting Angle on the Extraction Accuracy}
To test the effect of photographic perspective distortion on the accuracy of watermark extraction, photographs were taken at different horizontal angles.
The distance between the camera and the center of the screen was fixed at 40 cm.
The results of the experiment are listed in the Table \ref{angle_table}.

\begin{figure}[b!]
\begin{minipage}[h]{0.24\linewidth}
\center{\includegraphics[width=1\linewidth]{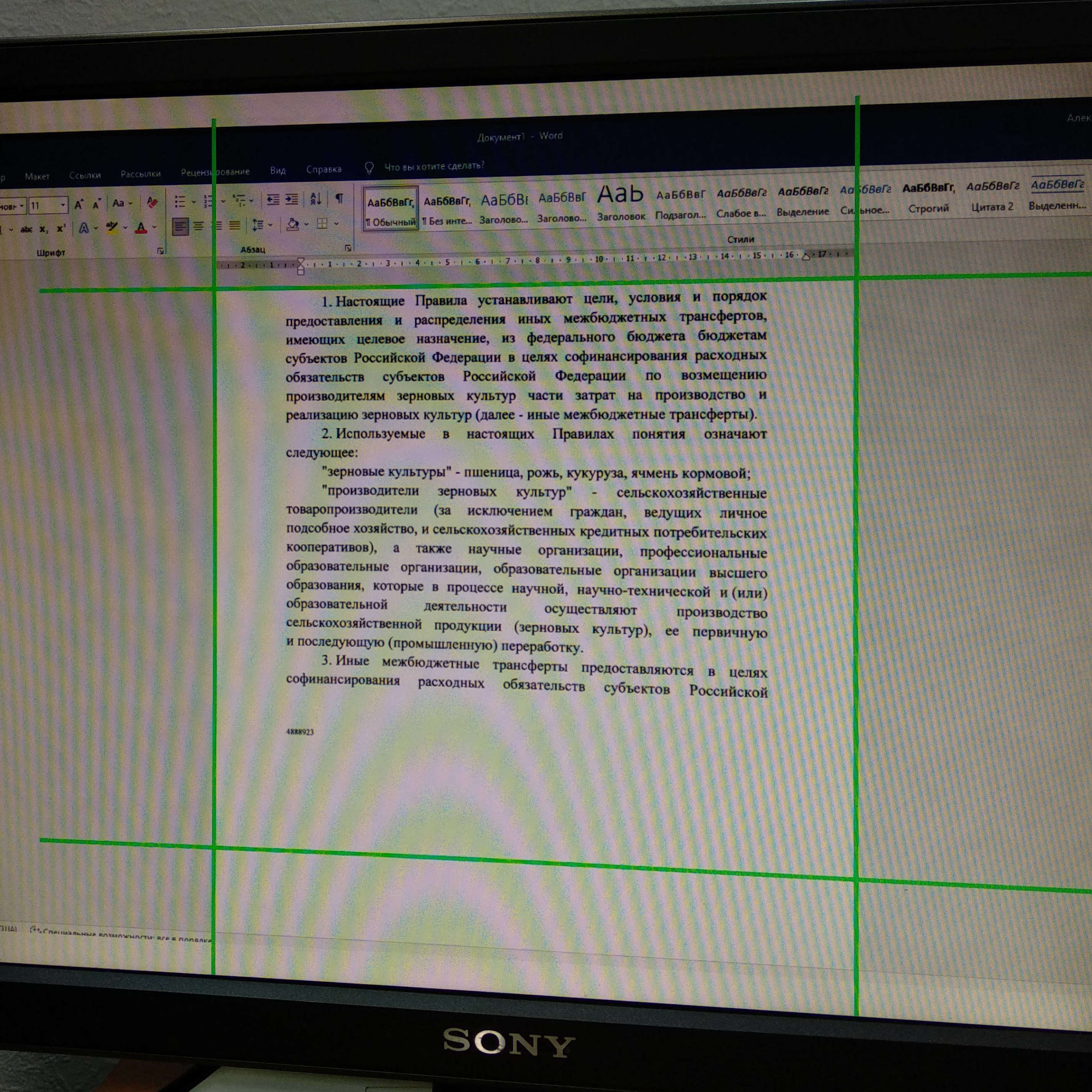}} \\ a) $15^\circ$
\end{minipage}
\hfill
\begin{minipage}[h]{0.24\linewidth}
\center{\includegraphics[width=1\linewidth]{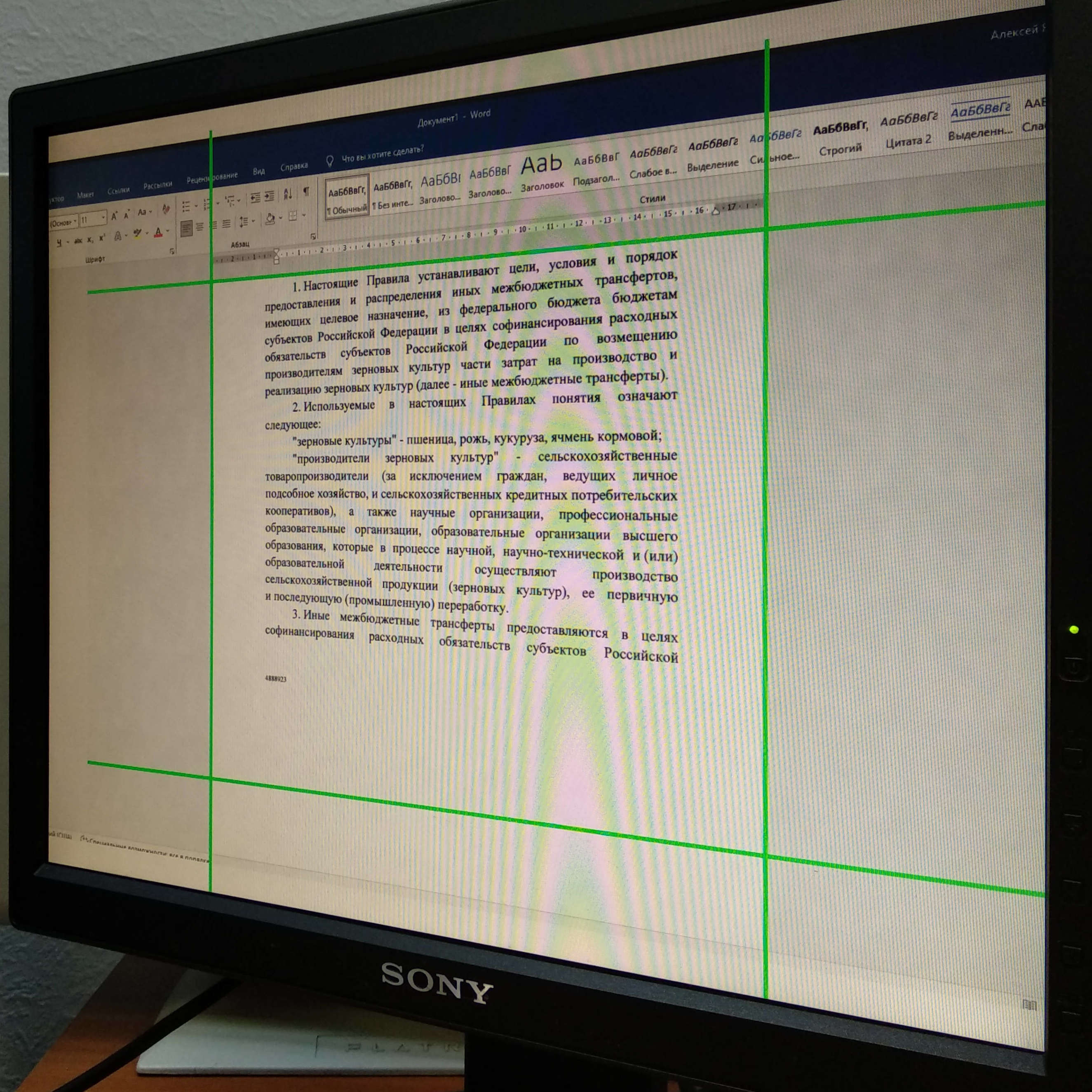}} \\ b) $30^\circ$
\end{minipage}
\hfill
\begin{minipage}[h]{0.24\linewidth}
\center{\includegraphics[width=1\linewidth]{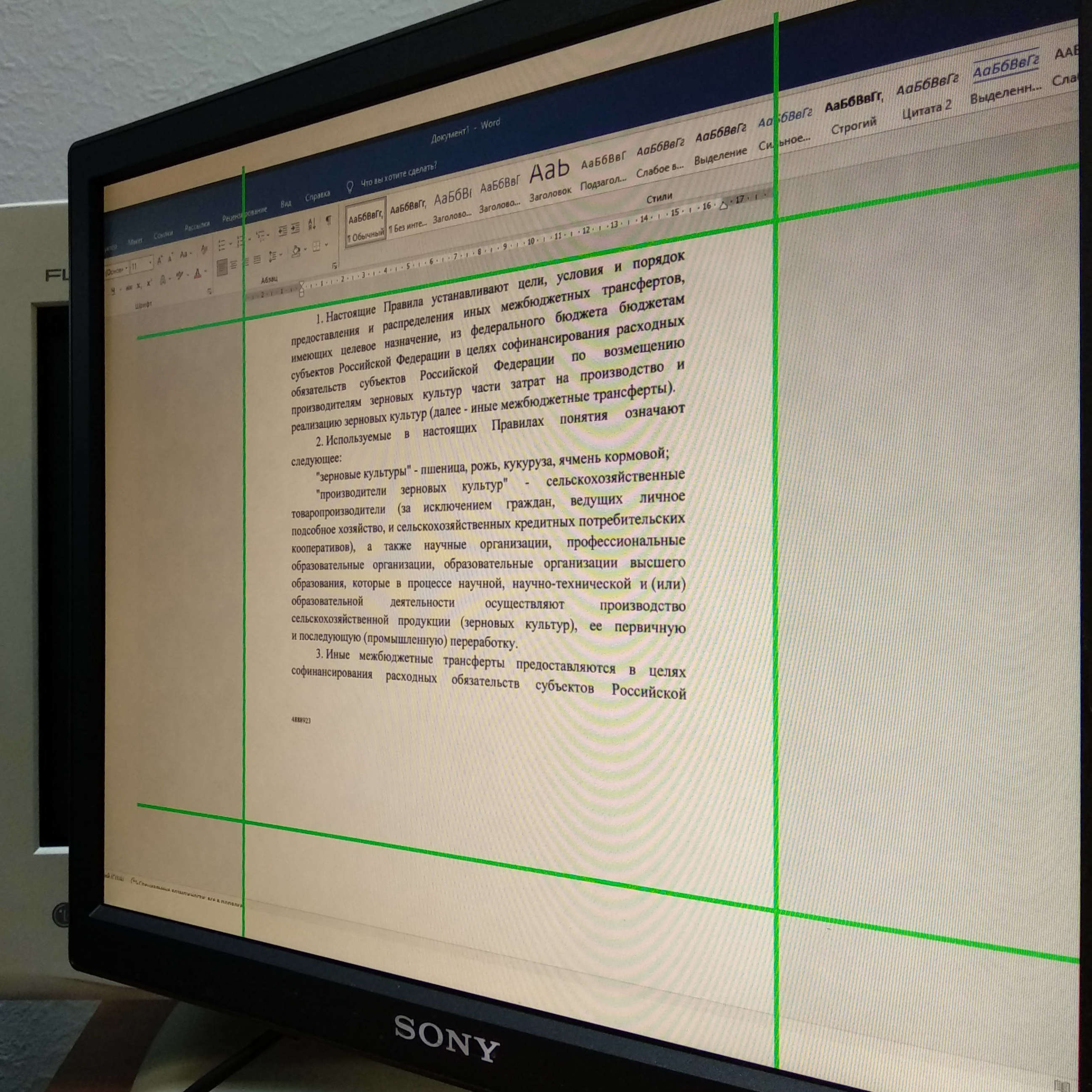}} \\ c) $45^\circ$
\end{minipage}
\hfill
\begin{minipage}[h]{0.24\linewidth}
\center{\includegraphics[width=1\linewidth]{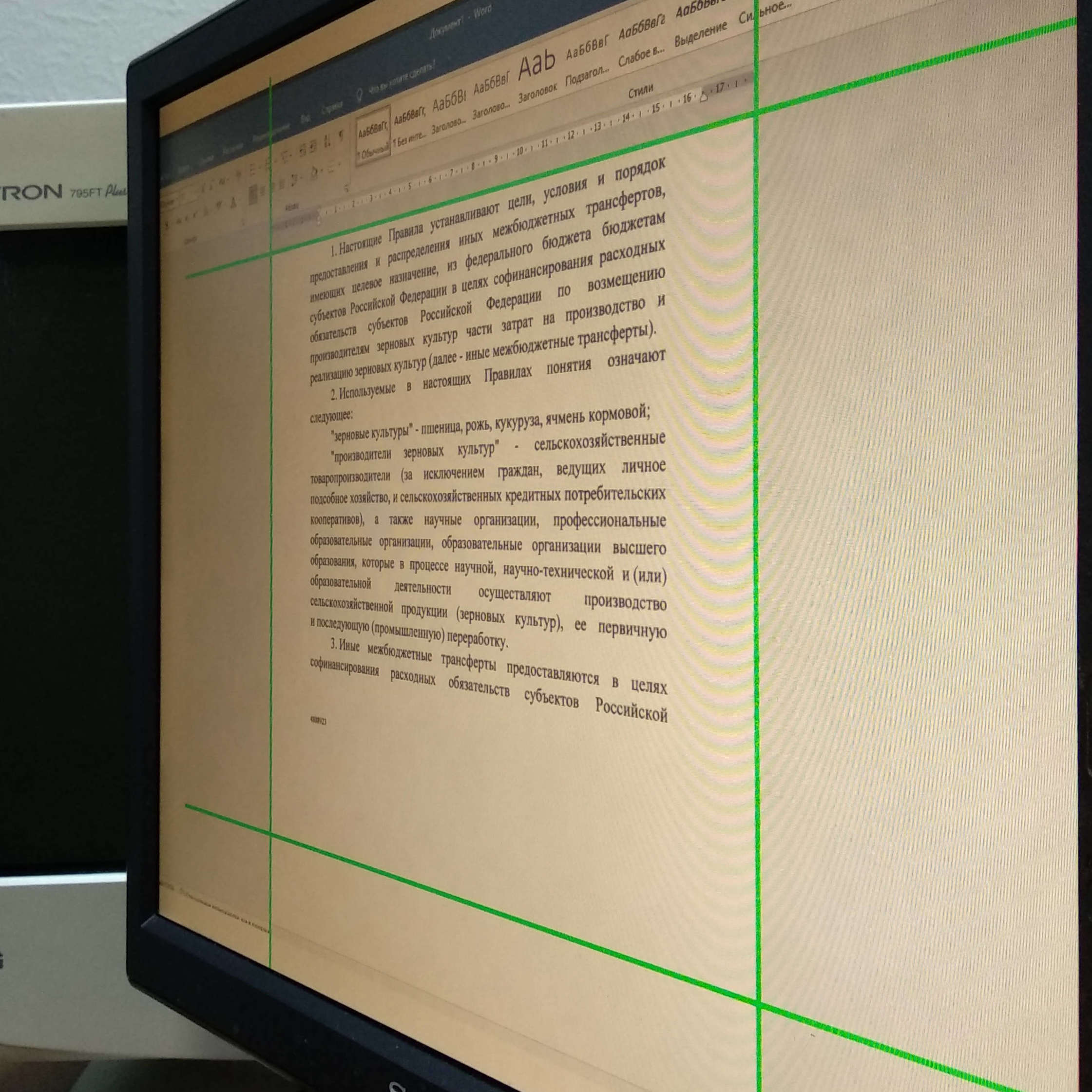}} \\ d) $60^\circ$
\end{minipage}
\caption{Photographs of the Sony SDM-S75A monitor at the distance of 40 cm with different perspective angles.}
\label{fig:angle_photos}
\end{figure}

\begin{table}[b!]
    \caption{The impact of perspective angle on the watermark extraction accuracy}
    \centering
    \begin{tabular}{c|cc|cc|cc}
    \hline
     & \multicolumn{6}{c}{Camera} \\ \cline{2-7}
    Angle & \multicolumn{2}{c|}{Xiaomi Mi A1} & \multicolumn{2}{c|}{Samsung S8} & \multicolumn{2}{c}{Samsung S21} \\ \cline {2-7}
    & BER & $\leqslant 3$ err. & BER & $\leqslant 3$ err. &BER & $\leqslant 3$ err.\\
    \hline
    \multicolumn{7}{c}{Samsung SyncMaster 940FN Monitor} \\
    \hline
    $0^\circ$ & 0.4\% & 10/10 & 0.2\% & 10/10 & 0.8\% & 10/10 \\
    $15^\circ$ & 0.8\% & 10/10 & 0.4\% & 10/10 & 1.2\% & 10/10 \\
    $30^\circ$ & 14.6\% & 6/10 & 1.0\% & 10/10 & 1.6\% & 9/10 \\
    $45^\circ$ & 0.4\% & 9/10 & 1.2\% & 10/10 & 0.2\% & 10/10 \\
    $60^\circ$ & 1.2\% & 10/10 & 1.4\% & 10/10 & 1.8\% & 10/10 \\
    \hline
    \multicolumn{7}{c}{Sony SDM-S75A Monitor} \\
    \hline
    $0^\circ$ & 32.0\% & 4/10 & 0.8\% & 10/10 & 22.0\% & 6/10 \\
    $15^\circ$ & 8.4\% & 8/10 & 1.6\% & 10/10 & 15.0\% & 7/10 \\
    $30^\circ$ & 2.0\% & 10/10 & 1.8\% & 10/10 & 1.4\% & 10/10 \\
    $45^\circ$ & 2.4\% & 10/10 & 1.4\% & 10/10 & 1.4\% & 10/10 \\
    $60^\circ$ & 1.4\% & 10/10 & 1.6\% & 10/10 & 1.0\% & 8/10 \\
    \hline
    \multicolumn{7}{c}{Dell U2722D Monitor} \\
    \hline
    $0^\circ$ & 1.0\% & 10/10 & 1.2\% & 10/10 & 0.8\% & 10/10 \\
    $15^\circ$ & 6.6\% & 8/10 & 7.4\% & 9/10 & 1.4\% & 10/10 \\
    $30^\circ$ & 1.4\% & 10/10 & 7.4\% & 9/10 & 1.8\% & 10/10 \\
    $45^\circ$ & 1.8\% & 10/10 & 3.8\% & 9/10 & 1.4\% & 10/10 \\
    $60^\circ$ & 1.6\% & 9/10 & 6.8\% & 9/10 & 1.2\% & 10/10 \\
    \hline
    \end{tabular}
    \label{angle_table}
\end{table}
Watermark extraction from photographs of the Sony SDM-S75A monitor taken with the cameras of Xiaomi Mi A1 and Samsung Galaxy S21 smartphones shows controversial result.
Photographs with a large perspective distortion are less prune to the moir\'{e} effect (Fig. \ref{fig:angle_photos}), which results in a higher watermark extraction accuracy.
The moir\'{e} effect also appears in the photographs of the Samsung SyncMaster 940FN monitor, taken with the camera of the Xiaomi Mi A1 smartphone at the angle of $15^\circ$, decreasing watermark extraction accuracy.

\subsection{The Impact of JPEG Compression Quality on the Extraction Accuracy}
The proposed document watermarking method was tested for robustness to JPEG image compression.
50 photographs with less than 3 extraction errors from previous experiments were selected randomly.
The photographs were compressed using the JPEG algorithm with different values of quality.
Compressed images extraction accuracy is presented in the Table \ref{jpeg_table}.

\begin{table}[t]
    \caption{The impact of JPEG compression quality on the watermark extraction accuracy}
    \centering
    \begin{tabular}{c|ccc}
    \hline
    JPEG & Average file & \multirow{2}{*} {BER} & \multirow{2}{*} {$\leqslant 3$ errors} \\
    quality & size, KB & &\\
    \hline
    Uncompressed & 4589 & 1.3\% & 50/50 \\
    80 & 1870 & 1.2\% & 50/50 \\
    60 & 1211 & 1.2\% & 50/50 \\
    50 & 1048 & 1.4\% & 49/50 \\
    40 & 900 & 1.4\% & 49/50 \\
    30 & 748 & 4.4\% & 46/50 \\
    20 & 573 & 8.5\% & 42/50 \\
    15 & 478 & 14.5\% & 34/50 \\
    10 & 376 & 26.9\% & 20/50 \\
    \hline
    \end{tabular}
    \label{jpeg_table}
\end{table}

The watermark is completely extracted from photographs after compression by the JPEG algorithm with at least 40 JPEG quality, and is also partially extracted with the quality at least 20.

\section{Conclusion}
Based on the existing approaches study and problem analysis, a new method for marking images of text documents displayed on a monitor screen has been developed.
A program for embedding a digital watermark on the monitor screen and a program for extracting a digital watermark from a screen photograph have been implemented.
The overlay image with the watermark is static and depends only on the encoded message and not on the screen image.
Due to the static nature of the method, the watermark embedding program consumes low computational resources.
Imperceptibility metrics have been calculated, according to which the watermark is considered invisible to the device user.
The proposed watermarking method was tested for the accuracy of extracting encoded message from a photograph for different pairs of camera and monitor, different distances from the camera to the screen, different angles between the camera and the screen.
Testing has shown high accuracy in extracting a message from screen photographs, with the exception of photographs taken in certain conditions.
The robustness of the watermarking method to the JPEG compression algorithm has been checked.
The main directions for further work are to increase of imperceptibility of the watermark on the screen and to improve the accuracy of watermark extraction from screen photographs.
The method shows low extraction accuracy from screen photographs with a strong moir\'{e} effect.
To solve this problem, it is necessary to study the possibility of applying existing methods for removing the moir\'{e} effect from screen photographs.
\newpage
\bibliographystyle{IEEEtran}
\IEEEtriggeratref{14}
\bibliography{main}

\begin{thebibliography}{10}
\providecommand{\url}[1]{#1}
\csname url@samestyle\endcsname
\providecommand{\newblock}{\relax}
\providecommand{\bibinfo}[2]{#2}
\providecommand{\BIBentrySTDinterwordspacing}{\spaceskip=0pt\relax}
\providecommand{\BIBentryALTinterwordstretchfactor}{4}
\providecommand{\BIBentryALTinterwordspacing}{\spaceskip=\fontdimen2\font plus
\BIBentryALTinterwordstretchfactor\fontdimen3\font minus
  \fontdimen4\font\relax}
\providecommand{\BIBforeignlanguage}[2]{{%
\expandafter\ifx\csname l@#1\endcsname\relax
\typeout{** WARNING: IEEEtran.bst: No hyphenation pattern has been}%
\typeout{** loaded for the language `#1'. Using the pattern for}%
\typeout{** the default language instead.}%
\else
\language=\csname l@#1\endcsname
\fi
#2}}
\providecommand{\BIBdecl}{\relax}
\BIBdecl

\bibitem{infowatch}
``\BIBforeignlanguage{russian}{Russia: restricted information leaks, 2020},''
  InfoWatch Analytics Center, 2021.

\bibitem{brassil1995electronic}
J.~T. Brassil, S.~Low, N.~F. Maxemchuk, and L.~O'Gorman, ``Electronic marking
  and identification techniques to discourage document copying,'' \emph{IEEE
  Journal on Selected Areas in Communications}, vol.~13, no.~8, pp. 1495--1504,
  1995.

\bibitem{cox1997secure}
I.~J. Cox, J.~Kilian, F.~T. Leighton, and T.~Shamoon, ``Secure spread spectrum
  watermarking for multimedia,'' \emph{IEEE transactions on image processing},
  vol.~6, no.~12, pp. 1673--1687, 1997.

\bibitem{hartung1999multimedia}
F.~Hartung and M.~Kutter, ``Multimedia watermarking techniques,''
  \emph{Proceedings of the IEEE}, vol.~87, no.~7, pp. 1079--1107, 1999.

\bibitem{premila_2008}
A.~Pramila, A.~Keskinarkaus, and T.~Sepp{\"a}nen, ``Multiple domain
  watermarking for print-scan and jpeg resilient data hiding,'' in
  \emph{Digital Watermarking}, 2008, pp. 279--293.

\bibitem{dong_2002}
P.~Dong and N.~Galatsanos, ``Affine transformation resistant watermarking based
  on image normalization,'' vol.~3, 2002, pp. 489--492.

\bibitem{ahvanooey_18}
M.~Taleby~Ahvanooey, Q.~Li, H.~J. Shim, and Y.~Huang, ``A comparative analysis
  of information hiding techniques for copyright protection of text
  documents,'' \emph{Security and Communication Networks}, vol. 2018, pp.
  1--22, 2018.

\bibitem{gslh}
D.~Gugelmann, D.~Sommer, V.~Lenders, M.~Happe, and L.~Vanbever,
  ``\BIBforeignlanguage{english}{Screen watermarking for data theft
  investigation and attribution},'' in \emph{\BIBforeignlanguage{english}{2018
  10th International Conference on Cyber Conflict (CyCon)}}.\hskip 1em plus
  0.5em minus 0.4em\relax IEEE, 2018, pp. 391--408.

\bibitem{our_gslh}
A.~Y. Yakushev, Y.~V. Markin, S.~A. Fomin, D.~O. Obydenkov, and B.~V.
  {kondrat'ev}, ``\BIBforeignlanguage{russian}{Text documents screen
  watermarking by changing background brightness in the interline spacing},''
  \emph{\BIBforeignlanguage{russian}{Proceedings of the Institute for System
  Programming of the RAS (Proceedings of ISP RAS)}}, vol.~33, no.~4, pp.
  147--162, 2021.

\bibitem{docmarking}
D.~O. Obydenkov, A.~Y. Yakushev, Y.~V. Markin, A.~E. Frolov, S.~A. Fomin, S.~V.
  Kozlov, D.~D. Gromey, A.~V. Kozachok, and B.~V. {Kondrat'ev}, ``Document
  marking system for leak investigations,'' \emph{Proceedings of the Institute
  for System Programming of the RAS (Proceedings of ISP RAS)}, vol.~33, no.~6,
  pp. 161--174, 2021.

\bibitem{dnn_screen}
S.~Ge, J.~Fei, Z.~Xia, Y.~Tong, J.~Weng, and J.~Liu, ``A screen-shooting
  resilient document image watermarking scheme using deep neural network,''
  \emph{IET Image Processing}, 2022.

\bibitem{fang2019camera}
H.~Fang, W.~Zhang, Z.~Ma, H.~Zhou, S.~Sun, H.~Cui, and N.~Yu, ``A camera
  shooting resilient watermarking scheme for underpainting documents,''
  \emph{IEEE Transactions on Circuits and Systems for Video Technology},
  vol.~30, no.~11, pp. 4075--4089, 2019.

\bibitem{fang2021tera}
H.~Fang, D.~Chen, F.~Wang, Z.~Ma, H.~Liu, W.~Zhou, W.~Zhang, and N.~Yu,
  ``{TERA}: Screen-to-camera image code with transparency, efficiency,
  robustness and adaptability,'' \emph{IEEE Transactions on Multimedia},
  vol.~24, pp. 955--967, 2021.

\bibitem{cheng2021mid}
Y.~Cheng, X.~Ji, L.~Wang, Q.~Pang, Y.-C. Chen, and W.~Xu, ``{mID}: Tracing
  screen photos via {Moir\'e} patterns,'' in \emph{30th USENIX Security
  Symposium (USENIX Security 21)}, 2021, pp. 2969--2986.

\bibitem{moire_challenge}
S.~Yuan, R.~Timofte, A.~Leonardis, and G.~Slabaugh,
  ``\BIBforeignlanguage{english}{Ntire 2020 challenge on image demoireing:
  Methods and results},'' in \emph{\BIBforeignlanguage{english}{Proceedings of
  the IEEE/CVF Conference on Computer Vision and Pattern Recognition
  Workshops}}, 2020, pp. 460--461.

\bibitem{screen_cam_moire}
Y.~Sun, Y.~Yu, and W.~Wang, ``\BIBforeignlanguage{english}{Moir{\'e} photo
  restoration using multiresolution convolutional neural networks},''
  \emph{\BIBforeignlanguage{english}{IEEE Transactions on Image Processing}},
  vol.~27, no.~8, pp. 4160--4172, 2018.

\bibitem{unet}
O.~Ronneberger, P.~Fischer, and T.~Brox, ``\BIBforeignlanguage{english}{U-net:
  Convolutional networks for biomedical image segmentation},'' in
  \emph{\BIBforeignlanguage{english}{International Conference on Medical image
  computing and computer-assisted intervention}}.\hskip 1em plus 0.5em minus
  0.4em\relax Springer, 2015, pp. 234--241.

\bibitem{circular_padding}
T.-H. Wang, H.-J. Huang, J.-T. Lin, C.-W. Hu, K.-H. Zeng, and M.~Sun,
  ``\BIBforeignlanguage{english}{Omnidirectional {CNN} for visual place
  recognition and navigation},'' in \emph{\BIBforeignlanguage{english}{2018
  IEEE International Conference on Robotics and Automation (ICRA)}}.\hskip 1em
  plus 0.5em minus 0.4em\relax IEEE, 2018, pp. 2341--2348.

\bibitem{EfficientNet}
M.~Tan and Q.~Le, ``\BIBforeignlanguage{english}{{EfficientNet}: rethinking
  model scaling for convolutional neural networks},'' in
  \emph{\BIBforeignlanguage{english}{International conference on machine
  learning}}.\hskip 1em plus 0.5em minus 0.4em\relax PMLR, 2019, pp.
  6105--6114.

\bibitem{ImageNet}
O.~Russakovsky, J.~Deng, H.~Su, J.~Krause, S.~Satheesh, S.~Ma, Z.~Huang,
  A.~Karpathy, A.~Khosla, M.~Bernstein \emph{et~al.},
  ``\BIBforeignlanguage{english}{{ImageNet} large scale visual recognition
  challenge},'' \emph{\BIBforeignlanguage{english}{International journal of
  computer vision}}, vol. 115, no.~3, pp. 211--252, 2015.

\bibitem{compositing}
T.~Porter and T.~Duff, ``\BIBforeignlanguage{english}{Compositing digital
  images},'' in \emph{\BIBforeignlanguage{english}{Proceedings of the 11th
  annual conference on Computer graphics and interactive techniques}}, 1984,
  pp. 253--259.

\bibitem{adam}
D.~P. Kingma and J.~Ba, ``\BIBforeignlanguage{english}{Adam: A method for
  stochastic optimization},'' \emph{\BIBforeignlanguage{english}{arXiv preprint
  arXiv:1412.6980}}, 2014.

\bibitem{bch}
R.~C. Bose and D.~K. Ray-Chaudhuri, ``\BIBforeignlanguage{english}{On a class
  of error correcting binary group codes},''
  \emph{\BIBforeignlanguage{english}{Information and control}}, vol.~3, no.~1,
  pp. 68--79, 1960.

\bibitem{minobr}
\BIBentryALTinterwordspacing
``\BIBforeignlanguage{russian}{List of documents on the website of the
  {Ministry of Science and Higher Education of the Russian Federation}}.''
  [Online]. Available: \url{https://minobrnauki.gov.ru/documents/}
\BIBentrySTDinterwordspacing

\bibitem{minfin}
\BIBentryALTinterwordspacing
``\BIBforeignlanguage{russian}{List of documents on the website of the
  {Ministry of Finance of the Russian Federation}}.'' [Online]. Available:
  \url{https://archive.minfin.gov.ru/en/}
\BIBentrySTDinterwordspacing

\bibitem{ssim}
Z.~Wang, A.~C. Bovik, H.~R. Sheikh, and E.~P. Simoncelli, ``Image quality
  assessment: from error measurement to structural similarity,'' \emph{IEEE
  transactions on image processing}, vol.~13, no.~1, 2004.

\bibitem{open_images}
A.~Kuznetsova, H.~Rom, N.~Alldrin, J.~Uijlings, I.~Krasin, J.~Pont-Tuset,
  S.~Kamali, S.~Popov, M.~Malloci, A.~Kolesnikov \emph{et~al.},
  ``\BIBforeignlanguage{english}{The open images dataset v4},''
  \emph{\BIBforeignlanguage{english}{International Journal of Computer
  Vision}}, vol. 128, no.~7, pp. 1956--1981, 2020.

\end{thebibliography}

\end{document}